\documentclass[a4paper]{article} 
\usepackage[english]{babel}

\usepackage{amsmath}	% Advanced maths commands
\usepackage{amssymb}	% Extra maths symbols
\usepackage{xcolor}
\usepackage{natbib}
\usepackage{authblk}
\usepackage{graphicx}
\usepackage{hyperref}

\newcommand{\be}{\begin{equation}}
\newcommand{\ee}{\end{equation}}

\newcommand{\ba}{\begin{eqnarray}}
\newcommand{\ea}{\end{eqnarray}}

\newcommand{\apj}{ApJ}
\newcommand{\apjs}{ApJS}
\newcommand{\apjl}{ApJ}
\newcommand{\mnras}{MNRAS}

\newcommand{\aap}{A\&A}

\newcommand{\apss}{Ap\&SS}

\newcommand{\changeurlcolor}[1]{\hypersetup{urlcolor=#1}}  

\newcommand*{\email}[1]{%
	\normalsize\href{mailto:#1}{#1}}

\DeclareSymbolFont{matha}{OML}{txmi}{m}{it}% txfonts
\DeclareMathSymbol{\varv}{\mathord}{matha}{118}

\title{The major role of eccentricity in the evolution of colliding pulsar-stellar winds}

\author[1]{Maxim V.Barkov}
\author[2]{Valenti Bosch-Ramon}

\affil[1]{Institute of Astronomy, Russian Academy of Sciences, Moscow, 119017 Russia; \email{barkov@inasan.ru}}
\affil[2]{ Departament de F\'{i}sica Qu\`antica i Astrof\'{i}sica, Institut de Ci\`encies del Cosmos (ICC), Universitat de Barcelona (IEEC-UB), Mart\'{i} i Franqu\`es 1, E08028 Barcelona, Spain; \email{vbosch@fqa.ub.edu}}
\begin{document}
%%%%%%%%%%%%%%%%%%%%%%%%%%%%%%%%%%%%%%%%%%
\maketitle

\begin{abstract}
{Binary systems that host a massive star and a non-accreting pulsar can be powerful non-thermal emitters. The relativistic pulsar wind and the non-relativistic stellar outflows interact along the orbit, producing ultrarelativistic particles that radiate from radio to gamma rays. To properly characterize the physics of these sources, and better understand their emission and impact on the environment, careful modelling of the outflow interactions, spanning a broad range of spatial and temporal scales, is needed. Full 3-dimensional approaches are very computationally expensive, but simpler approximate approaches, while still realistic at the semi-quantitative level, are available. We present here the results of calculations done with a quasi 3-dimensional scheme to compute the evolution of the interacting flows in a region spanning in size up to a thousand times the size of the binary. In particular, we analyze for the first time the role of different eccentricities in the large scale evolution of the shocked flows. We find that the higher the eccentricity, the closer the flows behave like a one-side outflow, which becomes rather collimated for eccentricity values $\gtrsim 0.75$. The simulations also unveil that the pulsar and the stellar winds become fully mixed within the grid for low eccentricity systems, presenting a more stochastic behavior at large scales than in the highly eccentric systems.}
\end{abstract}

\section{Introduction}\label{intro}

%%%%%%%%%%%%%%%%%%%%%%%%%%%%%%%%%%%%%%%%%%

Binary systems hosting a massive star and a non-accreting pulsar, or pulsar high-mass binaries (PHMB), can be powerful sources of gamma rays. The objects of this kind capable of gamma-ray emission pertain to the wider class of gamma-ray binaries, in which most of the non-stellar radiation is released in the gamma-ray energy range \citep[see, e.g.,][for these and related sources]{dub13,par19a,par19b}. The radiation is produced through the interaction of a relativistic pulsar wind and the outflows ejected by the star: Massive stars produce strong non-relativistic winds of supersonic nature and, in cases of very fast stellar rotation, quasi-Keplerian equatorial disks (or decretion disks) that flow outwards at subsonic speeds. These outflows interact with the pulsar wind and {later on interstellar matter} in a process in which ultrarelativistic particles are accelerated and produce emission from radio to gamma rays \citep[see, e.g.,][]{tav97,sie05,dub06,ner07,kha07,kon12,zab13,dub15,mol20,hub21b,2020ApJ...893L..39L,2021arXiv210609858K}.

The pulsar wind-stellar outflow interactions are complex, and different regions that can influence each other are relevant when trying to understand the evolution of the shocked flows. Even the associated radiation and its reprocessing can feedback on the flow dynamics, making the whole physical system highly non-linear. At small scales, there is the region right between the star and the pulsar in which flows are stopped and shocked by colliding against each other. After the collision, the flows become subsonic and hot, and start moving symmetrically sideways while pressure gradients lead to their reacceleration, getting supersonic again. This picture is roughly similar in both sides of the contact discontinuity separating the stellar and the pulsar shocked flows. Later on, the respective evolution of the flows largely differs due, for instance, to very different wind momentum rates and initial velocities, plus orbital effects.

Due to the large momentum rate of stellar winds, the stellar wind confines the pulsar wind, which bends over the pulsar. Shocked winds form an approximately axisymmetric curved structure that becomes conical further away from the binary. On those larger scales, if orbital motion were neglected, the shocked winds would move ballistically and form a conical shell made of shocked pulsar wind, surrounded by another shell of shocked stellar wind. The half-opening angle of the conical contact discontinuity would converge
%\footnote{In fact, in the absence of perturbations of any kind, the shocked flow speeds would eventually converge to their initial values.}
to a value that can be derived from the pulsar-to-stellar wind momentum rate ratio \citep[][]{bog08}:
\be
\eta = \frac{L_{sd}}{\dot{M}_w \varv_w c}\,,
\label{eq:eta}
\ee
where $L_{sd}$ is pulsar spin-down luminosity,  $\dot{M}_w$ and $\varv_w$ are the stellar mass-loss rate and wind speed, respectively.  Following \cite{1993ApJ...402..271E} and \cite{bog08}, the contact discontinuity of the cone-like structure has an approximate opening angle of
\be
\phi_c \approx \frac{\pi}{6}\left(4-\eta^{2/5}\right)\eta^{1/3}.
\label{eq:thc}
\ee

However, orbital motion is to be included in the colliding-wind picture, which makes a Coriolis force appear, a force that affects differently the pulsar and the stellar outflows due to the large relative velocity and density contrast. This differential Coriolis effect makes the stellar wind push on the shocked pulsar wind against the orbital rotation sense, creating a strong deflection of the interaction structure in that direction, and triggering a strong lateral shock in the shocked pulsar wind. As the two shocked flows have very different densities and velocities, they are prone to the occurrence of strong instabilities, such as Rayleigh-Taylor, Kelvin-Helmholtz, and Richtmyer-Meshkov, in the contact discontinuity \citep{bbp15}. Thus, as the flows move, the outflow contact surface gets partially disrupted, stellar wind mixes with the shocked pulsar wind, and the latter develops strong turbulence and decelerates. The result is that the shocked flow structure shape becomes a one-arm spiral that fills much of the volume and is expected to disrupt after a few turns.

Between the apex of the interaction structure, located at the two-shocked flow stagnation point \citep[][]{bog12}, and the starting point of the Coriolis shock, on the leading edge of the interaction structure, the shocked pulsar wind gets compressed and thus heated by the Coriolis force-related lateral pressure of the stellar wind. This can weaken the mentioned shocked flow reacceleration caused by pressure gradients. On the other hand, in the trailing edge of that interaction structure, the shocked pulsar wind quickly expands and accelerates through rarefaction waves. 

The presence of decretion disks can significantly alter the geometry of the interaction structure, which must develop now embedded in a much more complex circumstellar environment. Nevertheless, the disk is rather massive and marginally bounded to the star, so part of the material may not even escape the binary. In addition, the accumulated disk mass can be in fact just comparable to that of the stellar wind. Thus, on large scales, the shocked flow dynamics is likely dominated by the pulsar and the stellar wind, the disk  and radiation processes can be important for flow dynamics on small and middle scales. Assuming then that the disk is mostly relevant closer to the binary, and neglecting the role of the magnetic field, the system eccentricity may turn out to be as important as $\eta$ to describe the evolution of the shocked flows on large scales. In particular, \cite{bbmb17,2018MNRAS.479.1320B} show that for very high eccentricities the shocked pulsar wind becomes strongly focused along the periastron-apastron direction, as it gets deflected by the stellar wind in that direction for most of the orbit. To date, however, an exploration of how different eccentricity values affect the large-scale shocked flow structure is missing, mainly, at which eccentricities one-sided outflows form.

In this work, we perform a numerical study of how the shocked flows from PHMB evolve, and propagate, up to large distances from the binary for different eccentricities. Our major goal is to find for which orbit eccentricity the mentioned one-sided outflow forms. The study is carried out using the quasi 3-dimensional (3D) calculation scheme developed by \cite{bb16}, used to study PSR~B1259$-$63, and HESS~J0632$+$057 \citep{bbmb17,2018MNRAS.479.1320B}. The advantage of this method, which employs spherical coordinates, is that it focuses on the orbital plane, but sacrifices resolution for zenital angles far from that plane. Furthermore, the shocked flow geometry just outside the binary turns out to be amenable to be simplified such that the colliding-wind apex region does not need to be modelled, which allows a computationally much cheaper resolution. All this largely reduces the cost of the simulations, allowing one to probe a large region surrounding the binary. The accuracy of the method is appropriate at a semi-quantitative level, as shown by comparison with results obtained using full 3D calculations encompassing overlapping regions \citep{bbp15}.

The article is organized as follows: In Sect.~\ref{model}, the simulations are described, and in Sect.~\ref{res}, their results presented. Then, in Sect~\ref{dis}, the simulations results are summarized and discussed.

\section{Numerical model}\label{model}

Quasi-3D simulations of PHMB wind-wind collisions with different orbit eccentricities were performed using the {\it PLUTO} code\footnote{Link http://plutocode.ph.unito.it/index.html} \citep{mbm07}. {\it PLUTO} is a modular Godunov-type code entirely written in C and intended mainly for astrophysical applications and high Mach number flows in multiple spatial dimensions. Spatial parabolic interpolation, a 3rd order Runge-Kutta approximation in time, and an HLLC Riemann solver were used \citep{2005JCoPh.203..344L}. The simulations were performed on the CFCA XC30 cluster of the National Astronomical Observatory of Japan (NAOJ). To reduce computing costs, the flow was approximated by a simple equation of state enough for our purposes: that of an ideal relativistic gas with adiabatic index $4/3$. We adopted spherical coordinates $(R,\theta,\phi)$, with 768 cells in both the radial and the azimuthal directions. To reduce the computation costs, we took only 3 cells in the zenital direction. The domain size was taken to be $R \in [1, 500]R_{min}$, $\theta \in [\pi/4, 3\pi/4]$ and $\phi \in [0, 2\pi]$. We set $R_{min} = 2(1-e^2)a$, with $a$ being the semi-major axis of the orbit, and $e$ its eccentricity; thus, the scales captured by the simulations are larger than a few times the orbital separation distance at periastron. The computational grid was made logarithmic in the radial direction, that is, cells grow with a constant aspect ratio. Our treatment of the $\theta$-direction allows a reasonably realistic characterization of the orbital plane physics on scales beyond the pulsar, where the interaction structure expansion becomes approximately linear with distance, although a more quantitative account would require a complete 3D treatment \citep[see][]{bb16,bbmb17,2018MNRAS.479.1320B}.

In the quasi-3D calculation scheme adopted here the injected pulsar wind has a half-opening angle $\phi_c$, which depends on the momentum rate relation, which is set to $\eta = 0.1$, an intermediate value for this parameter, to restrict the degrees of freedom of the problem. This $\eta$-value corresponds to $\phi_c = 0.87$~radians, so the computational domain was divided in two non-equal parts. The first one, in which $\phi \in (\phi_c, 2\pi-\phi_c)$, was filled by a radial stellar wind with velocity value $\varv_w=2400$~km~s$^{-1}$. The second part, in which $\phi \in [-\phi_c,\phi_c]$, was filled by a radial pulsar wind with Lorentz factor $\Gamma=1.9$. Despite this Lorentz factor being just moderately relativistic, the calculations are enough relativistic for our purposes { because internal energy density already plays a significant inertial role} \citep[see][]{bbkp12,bbp15}. The winds were assumed to be highly supersonic at injection, with Mach numbers $M_w=6.9$ and $M_j = \gamma_p \varv_p/ \gamma_{s,p} c_{s,p} = 14$ for the stellar and the pulsar wind, respectively, where "s" refers to the sound speed. For simplicity, any wind azimuthal velocity component, coming for instance from object rotation and angular momentum conservation, was neglected in our calculations as its value would be well below $\varv_w$ and $c$. Also, as discussed in Sect.\ref{intro}, we neglected the role of a decretion disk, although more quantitative studies should include it.

The initial pulsar position is at the left of the computational domain, which means that the simulated pulsar wind cone is also initially directed to the left, which corresponds to the periastron-apastron direction. The adopted orbital period is $T_{orb} = 16.6$~days, and the stellar masses are $M_{MS} = 31$~M$_\odot$ and $M_{PSR} = 1.44$~M$_\odot$ for the star and the pulsar, respectively, {\ so from Kepler's third law} the corresponding orbital semi-major axis is $a=6\times10^{12}$~cm. During the simulation, the $\phi$-intervals within which the pulsar and the stellar winds are injected rotate along the orbit with the pace and sense of the corresponding orbital velocity. The studied cases have orbit eccentricities $e=0$, 0.25, 0.5 and 0.75. Simulations with even larger $e$-values can be found in \cite{bb16,bbmb17,2018MNRAS.479.1320B}. These $e$-values are characteristic of the different known high-mass gamma-ray binaries, although we note that a pulsar has been confirmed to be present only in PSR~B1259$-$63 and PSR~J2032+4127 \citep[e.g.][]{aha05,lyn15}. On the other hand, the simulated orbital period is significantly shorter than those of HESS~J0632$+$057, PSR~B1259$-$63 and PSR~J2032+4127. {This was done as previous studies already explored cases with long periods \citep{bb16,bbmb17,2018MNRAS.479.1320B}}. It is worth noting that the asymmetry of the interaction structure on large scales should not depend significantly on $T_{orb}$, because what characterizes this asymmetry is the relative change with $\phi$ of the shocked pulsar wind energy rate along the orbit, which is independent of $T_{orb}$. We note that a potential $\varv_\phi$ component of the stellar wind would affect the angular distribution of the flow, but since $\varv_\phi\propto 1/R$ this component will become negligible at $R\gg R_{min}$.

%%%%%%%%%%%%%%%%%%%%%%%%%%%%%%%%%%%%%%%%%%
\section{Results}\label{res}

\begin{figure}
%\widefigure
\includegraphics[width=6.7 cm]{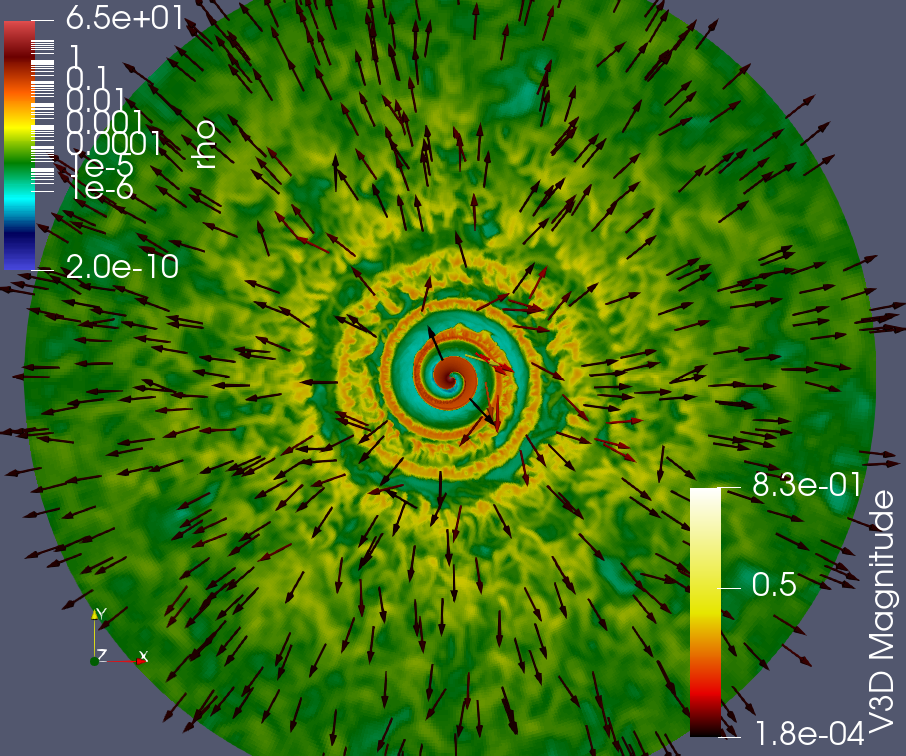}
\includegraphics[width=6.7 cm]{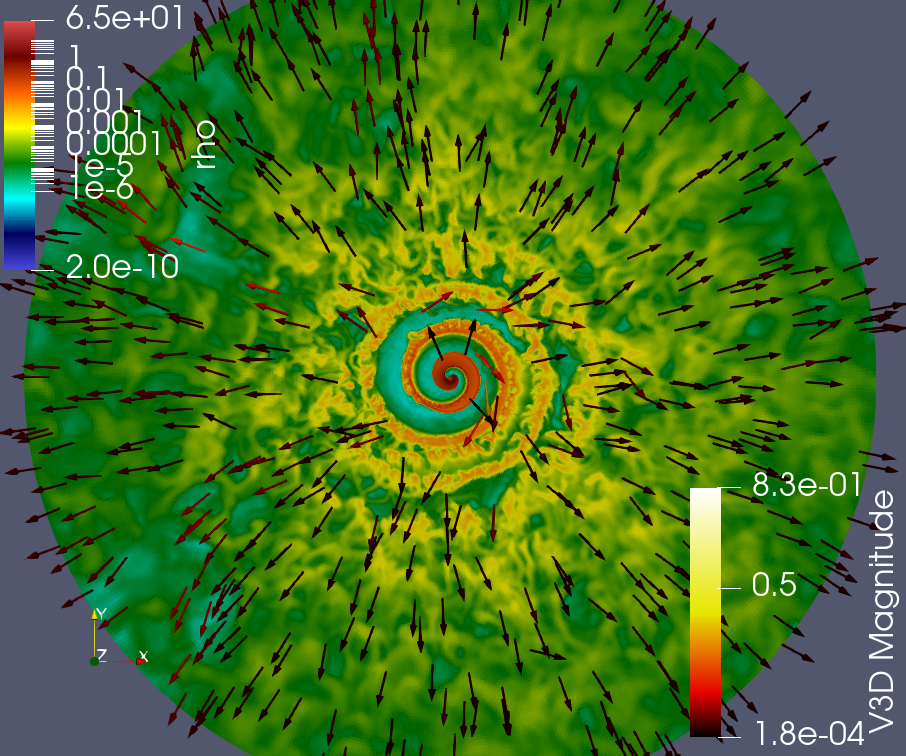}
\includegraphics[width=6.7 cm]{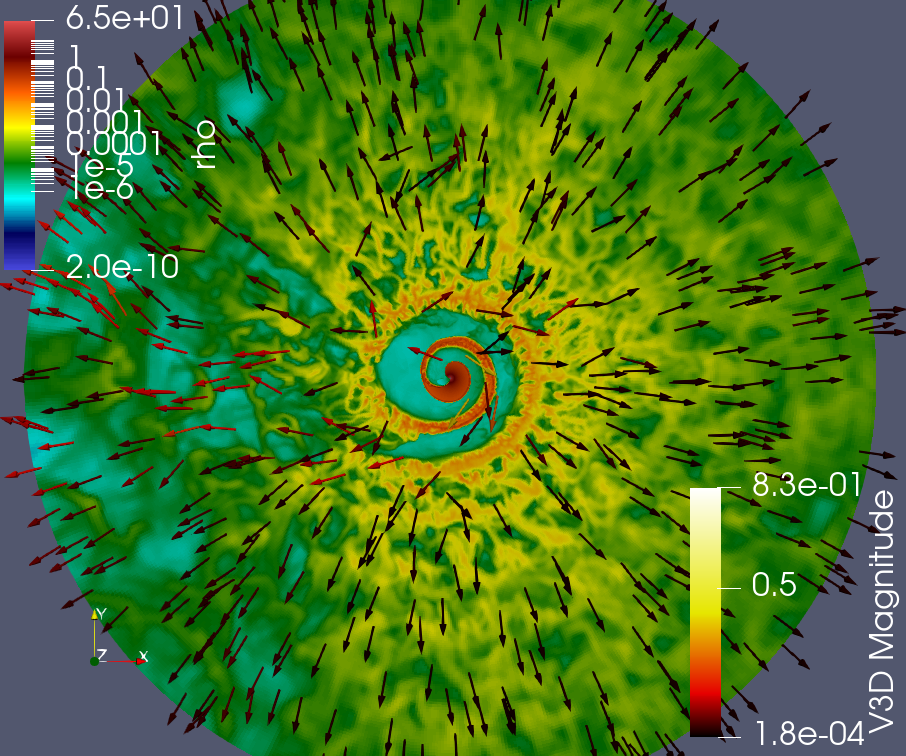}
\includegraphics[width=6.7 cm]{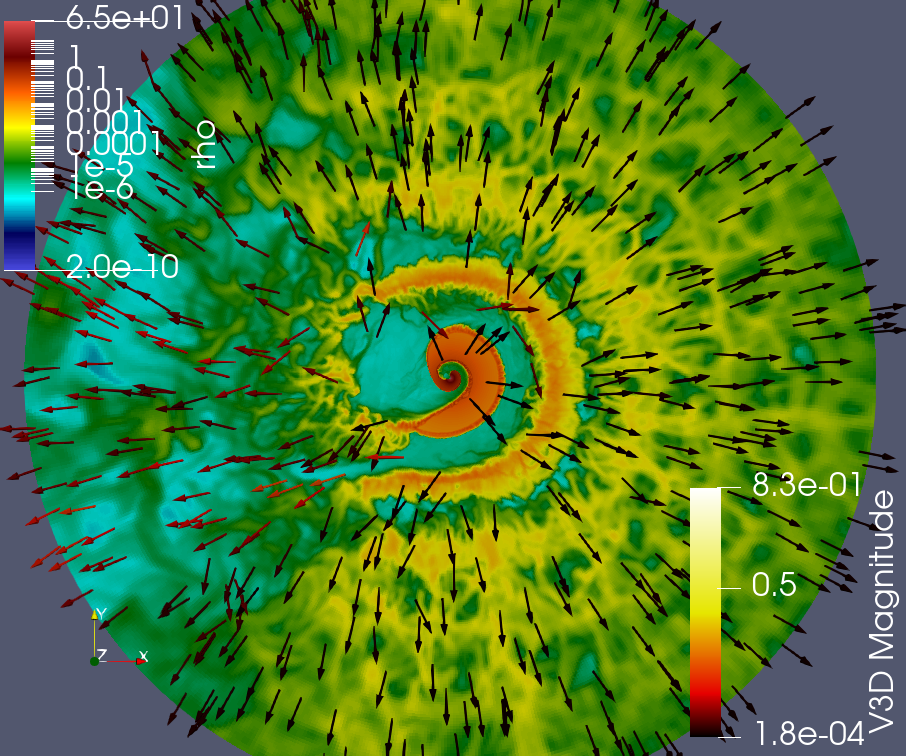}
\caption{{Maps for the whole computational domain} of the logarithm of the density distribution on the orbital plane by color, with colored arrows representing the velocity field, for different eccentricities: $e=0$ (top-left panel); $e=0.25$ (top-right panel); $e=0.5$ (bottom-left panel); and $e=0.75$ (bottom-right panel). The radius of the circular grid region shown is $1000(1-e^2)a$, and units of the legend scales are $a=6\times10^{12}$~cm. {The periastron-apastron direction is to the left.}}
\label{fig:rhov}
\end{figure}

\begin{figure}	
%\widefigure
\includegraphics[width=6.7 cm]{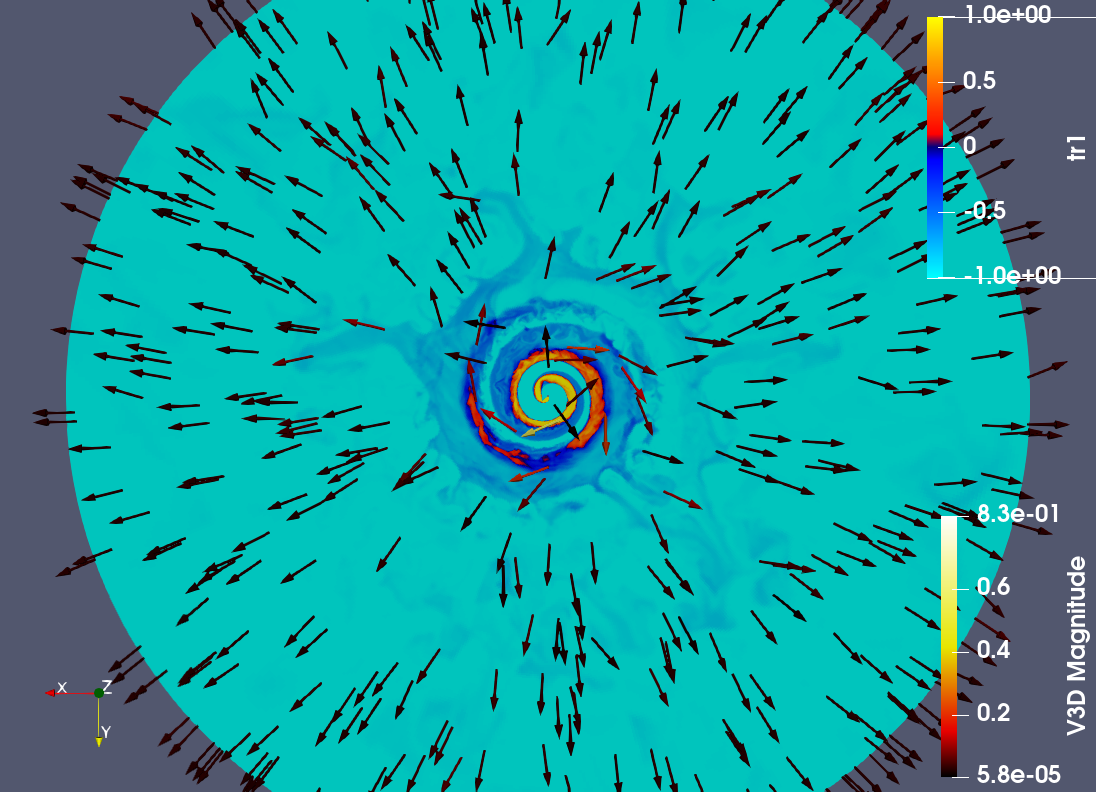}
\includegraphics[width=6.7 cm]{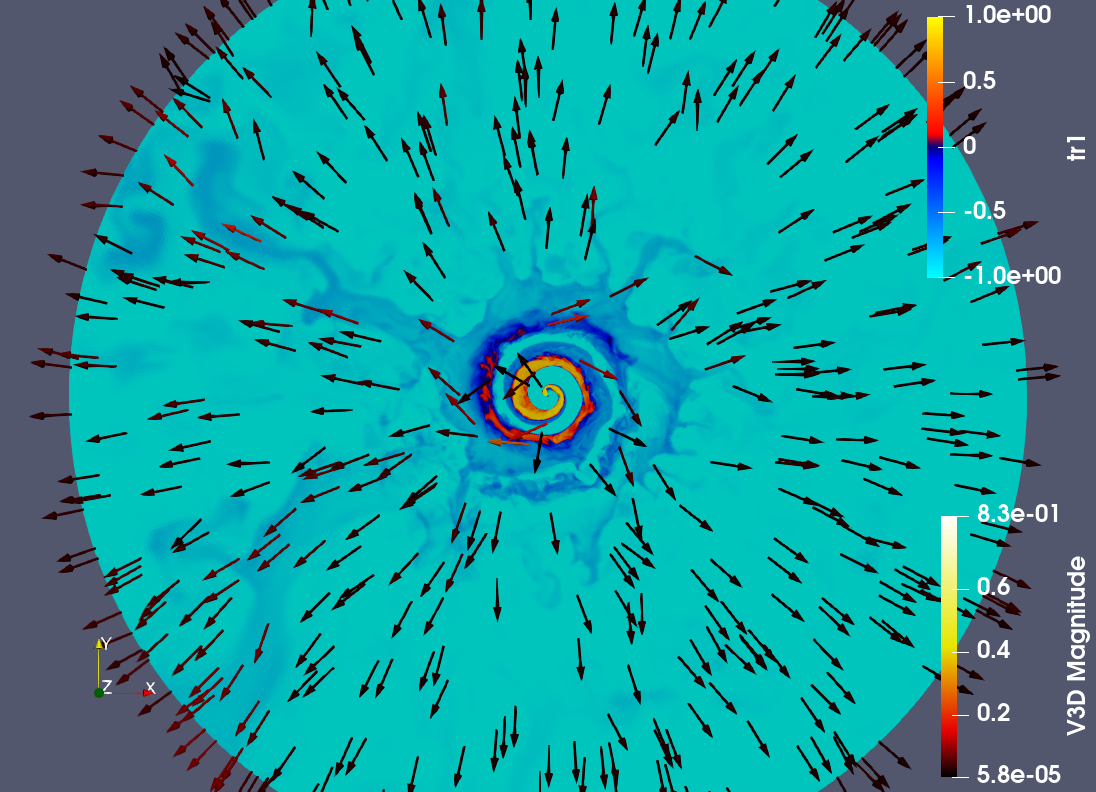}
\includegraphics[width=6.7 cm]{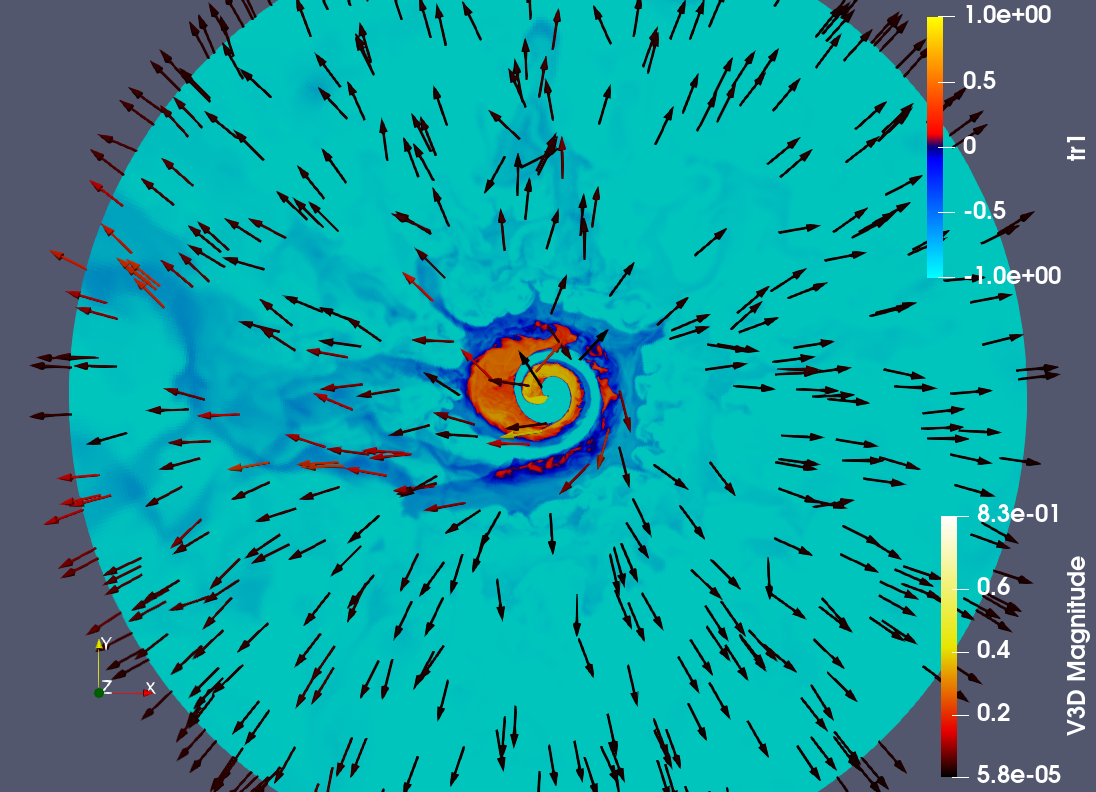}
\includegraphics[width=6.7 cm]{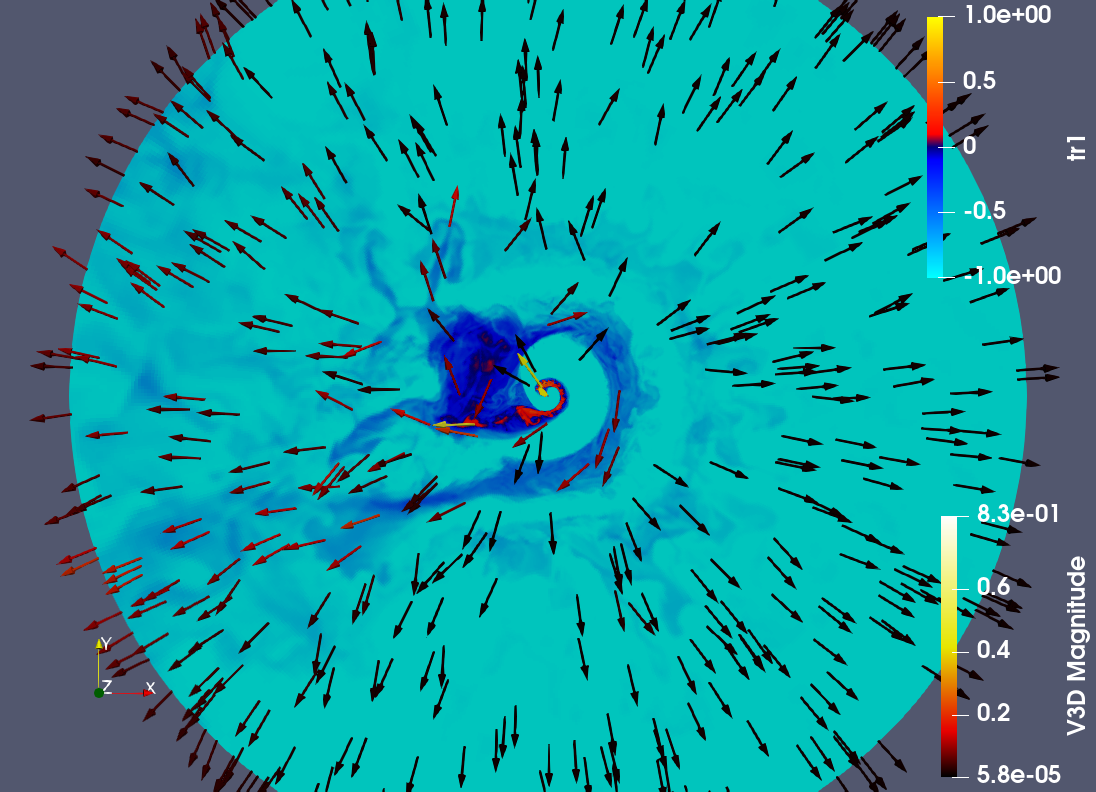}
\caption{The same as in Fig.~\ref{fig:rhov} but for tracer, where 1 and -1 correspond to the injected pulsar and stellar winds, respectively.}
\label{fig:tr1v}
\end{figure}

\begin{figure}	
%\widefigure
\includegraphics[width=6.7 cm]{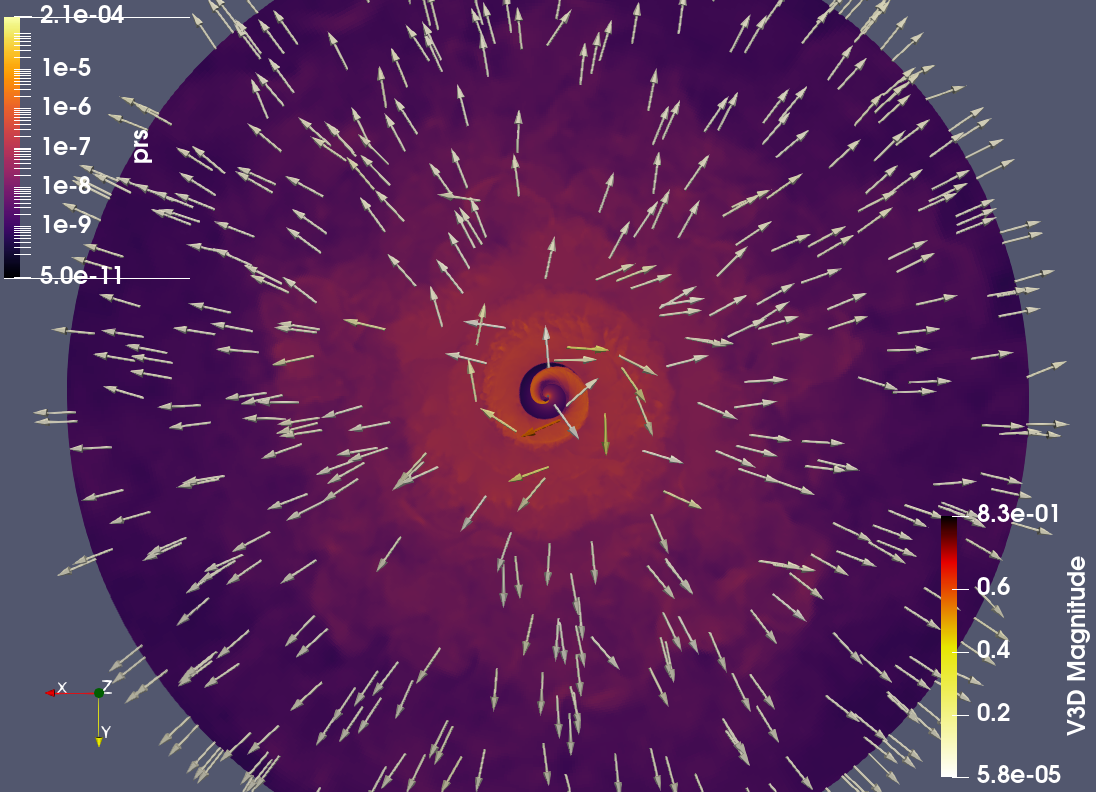}
\includegraphics[width=6.7 cm]{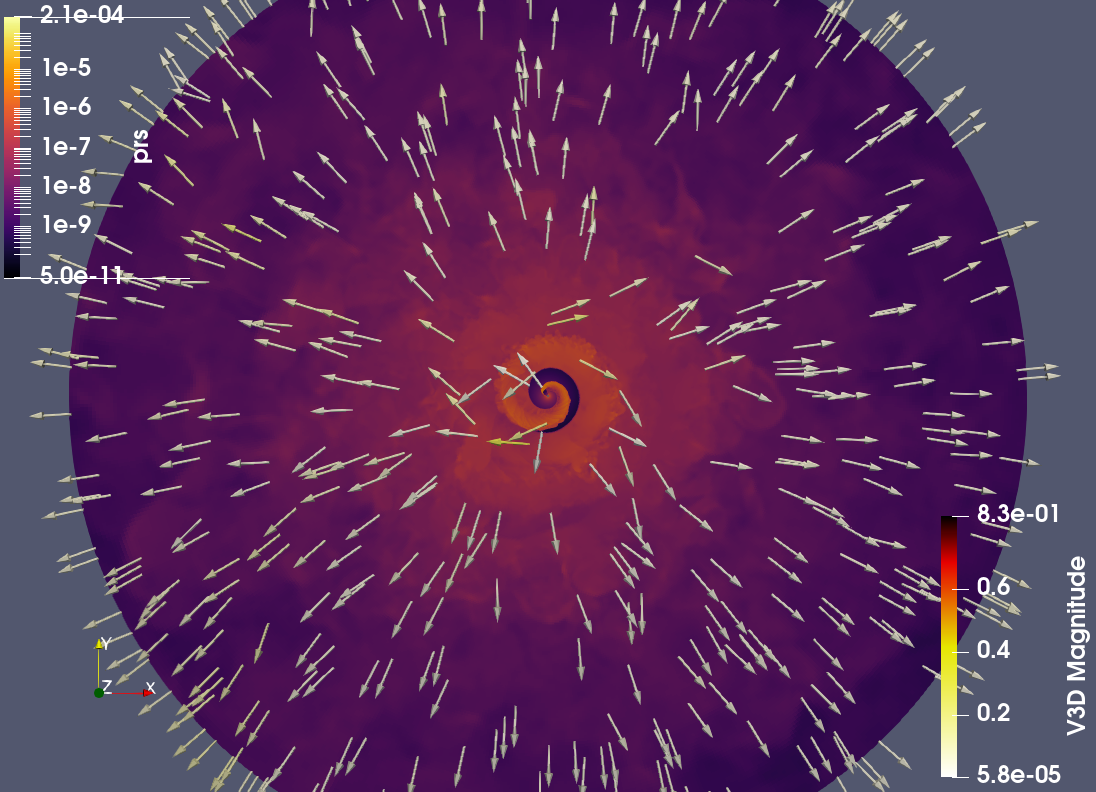}
\includegraphics[width=6.7 cm]{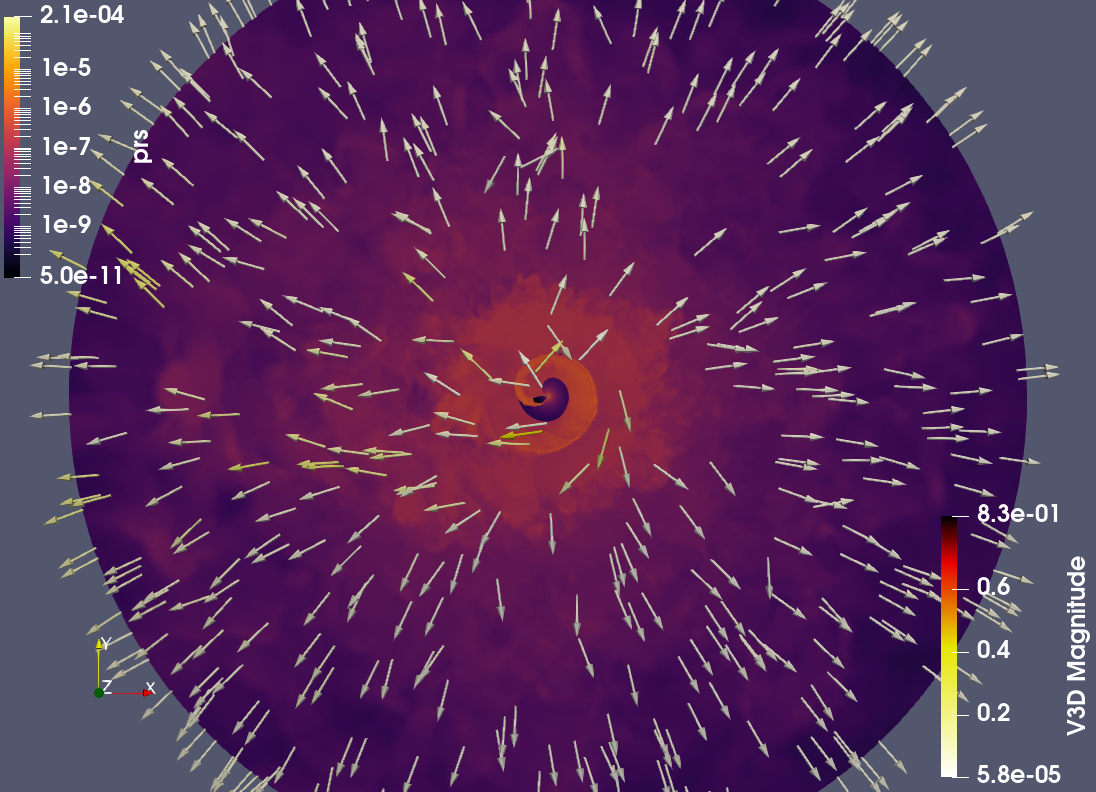}
\includegraphics[width=6.7 cm]{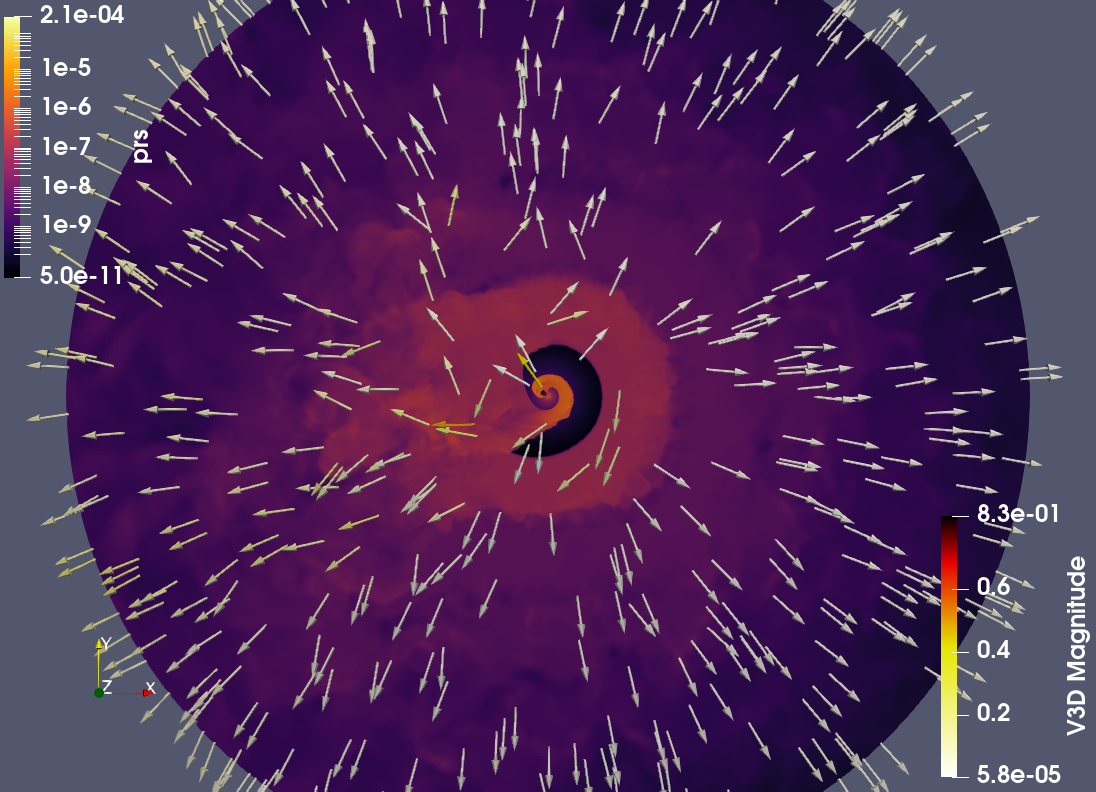}
\caption{The same as in Fig.~\ref{fig:rhov} but for pressure.}
\label{fig:prsv}
\end{figure}

Maps of the density distributions and the velocity vector fields in the orbital plane are presented in Fig.~\ref{fig:rhov} for the different eccentricities studied. Regardless of the eccentricity, the spiral structure starts to disrupt after one orbital turn, and after 2-3 orbital turns, spiral structure disruption leaves a more or less uniform medium with randomly located density and velocity irregularities. In particular, the velocity field shows spiral motion in the first turns and, farther away from the system, it mostly shows a radial outflow. Asymmetry of the interaction structure on the orbital plane starts to become significant  for $e\gtrsim 0.5$. In the case of $e=0.75$, the effect becomes extreme, with the density (velocity) in the periastron-apastron direction being much smaller (larger) than in the other directions. For $e=0.5$, this effect is also present, but not so strong. A fluid property used to track fluid motion  -- the tracer behavior, shown in Fig.~\ref{fig:tr1v}, is similar to that of the density, and one can see there as well that the spiral structure disappears after 2-3 orbital turns regardless of eccentricity. The higher the $e$-value, the more prominent the pulsar wind becomes in the periastron-apastron direction. The pressure spatial distribution is presented in Fig.~\ref{fig:prsv}, showing the same trends as density and tracer. In addition, pressure shows a smooth drop after spiral structure disruption. As for density and tracer, the main difference between cases with low and high eccentricities is that pressure falls anisotropically, its decrease being slightly shallower in the periastron-apastron direction when $e$ is large enough. 

We averaged, weighting in mass, the radial velocity and the Mach number in three $\phi$-ranges, or sectors: s1, with $\phi_{s1} \in [150^o, 210^o]$; s2, with $\phi_{s2} \in [290^o, 350^o]$; and s3, with $\phi_{s3} \in [60^o, 120^o]$. We also performed the mass-weighted average of the radial velocity in $R$, where $R\in [350\,a,435\,a]$. The mass-weighted value, averaged over a certain $R$- or $\phi$-range, was calculated as
\be
<q(\phi)> = \frac{ \int_{R \in [R1,R2]} q(R,\phi)\rho(R,\phi) dR}{\int_{R \in [R1,R2]} \rho(R,\phi) dR }
\label{eq:masev1}
\ee
and
\be
<q(R)> = \frac{ \int_{\phi \in sX} q(R,\phi)\rho(r,\phi) d\phi}{\int_{\phi \in sX} \rho(R,\phi) d\phi }\,, 
\label{eq:masev2}
\ee
respectively, where $\rho$ is density, $q$ the value being averaged, and $sX$ the corresponding sector. 

\begin{figure}	
%\widefigure
\includegraphics[width=6.7 cm]{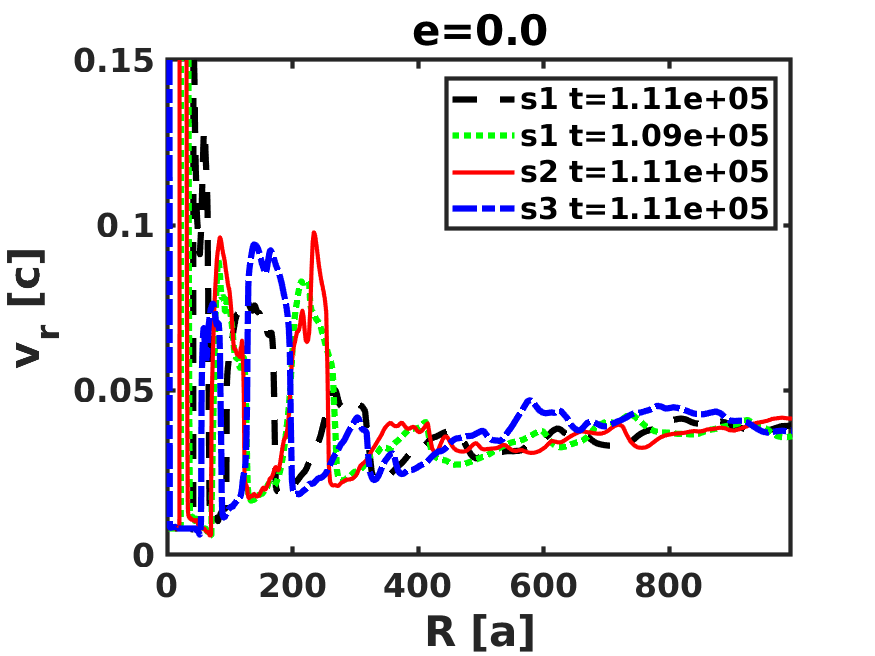}
\includegraphics[width=6.7 cm]{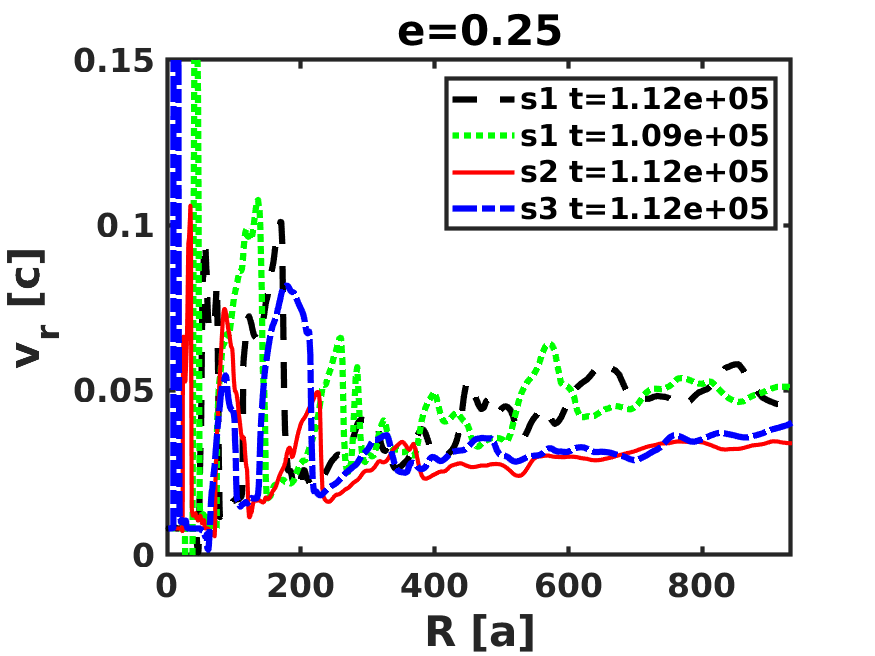}
\includegraphics[width=6.7 cm]{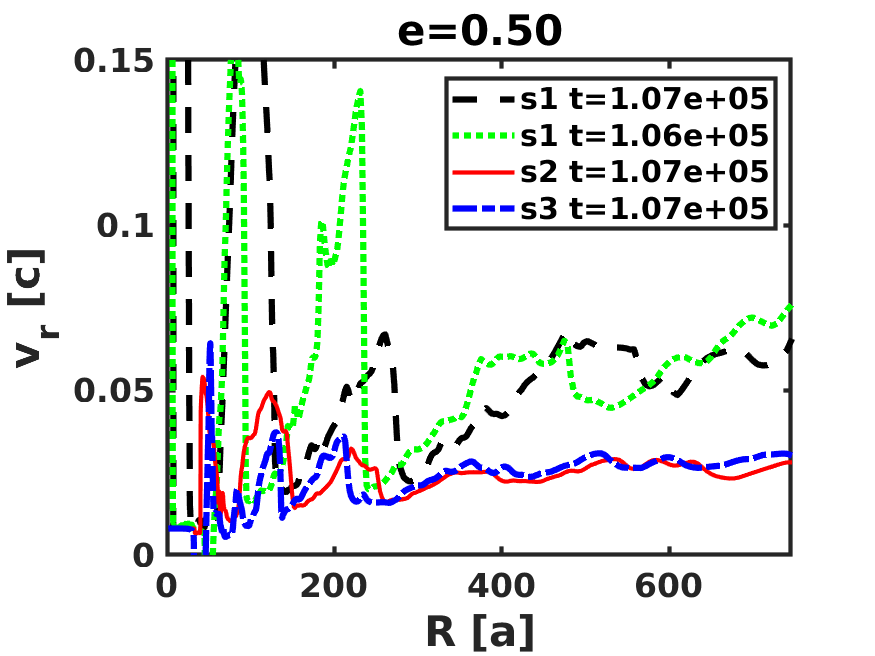}
\includegraphics[width=6.7 cm]{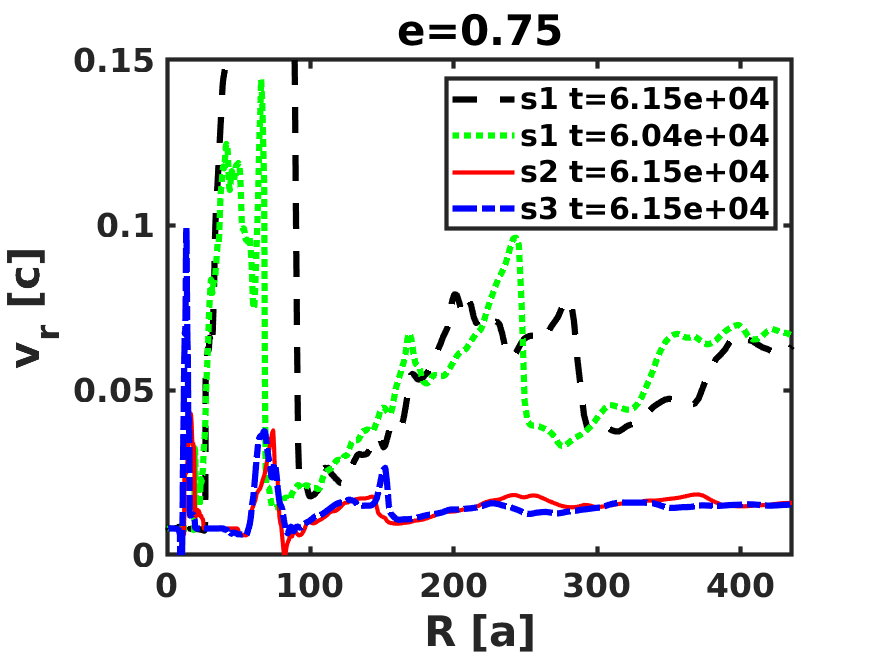}
\caption{Radial velocity, averaged over sectors s1, s2 and s3, versus $R$ %at $T=1.11\times 10^5$ 
and different orbital eccentricities: $e=0$ (left top panel); $e=0.25$ (right-top panel); $e=0.5$ (left-bottom panel); and $e=0.75$ (right-bottom panel). An additional curve for the s1 case at a slightly earlier time, %$T=1.09\times 10^5$, 
is also shown to illustrate the quick variability of the flow properties on top of its longer term behavior.}
\label{fig:vavR}
\end{figure}

\begin{figure}	
%\widefigure
\includegraphics[width=6.7 cm]{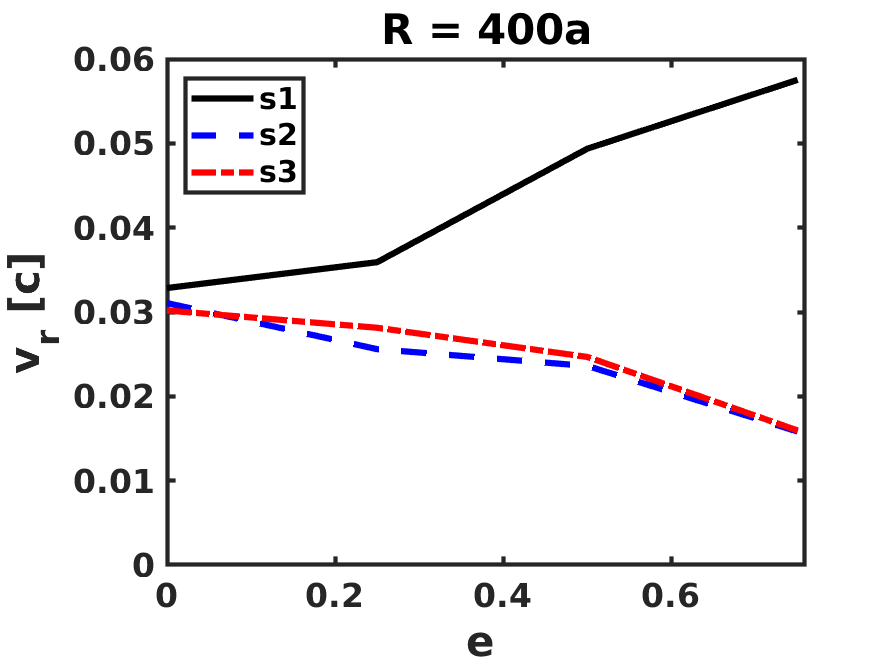}
\caption{Radial velocity, averaged over sectors s1, s2 and s3 at $R=400\,a$, versus orbital eccentricity.}
\label{fig:vave}
\end{figure}

\begin{figure}	
%\widefigure
\includegraphics[width=6.7 cm]{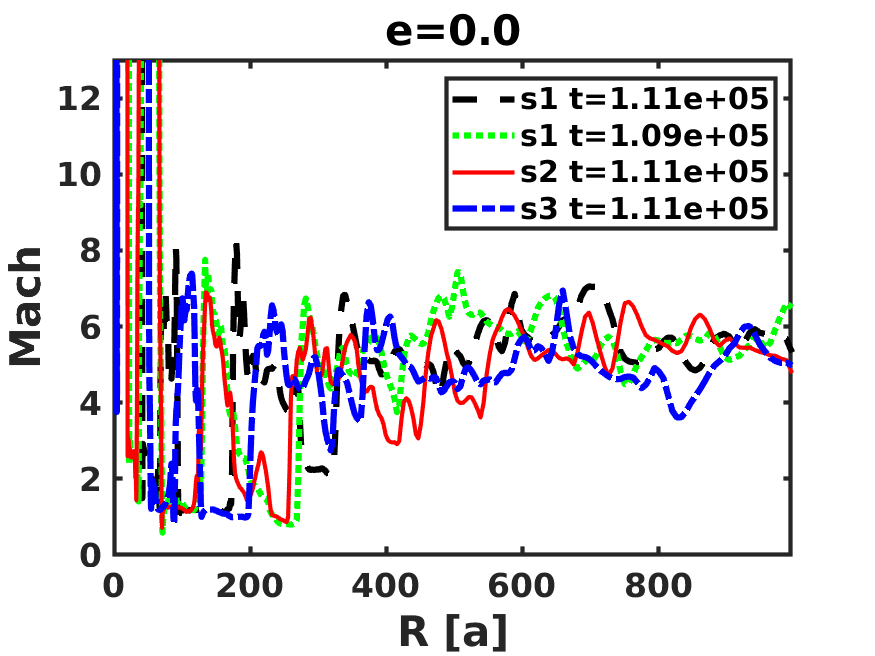}
\includegraphics[width=6.7 cm]{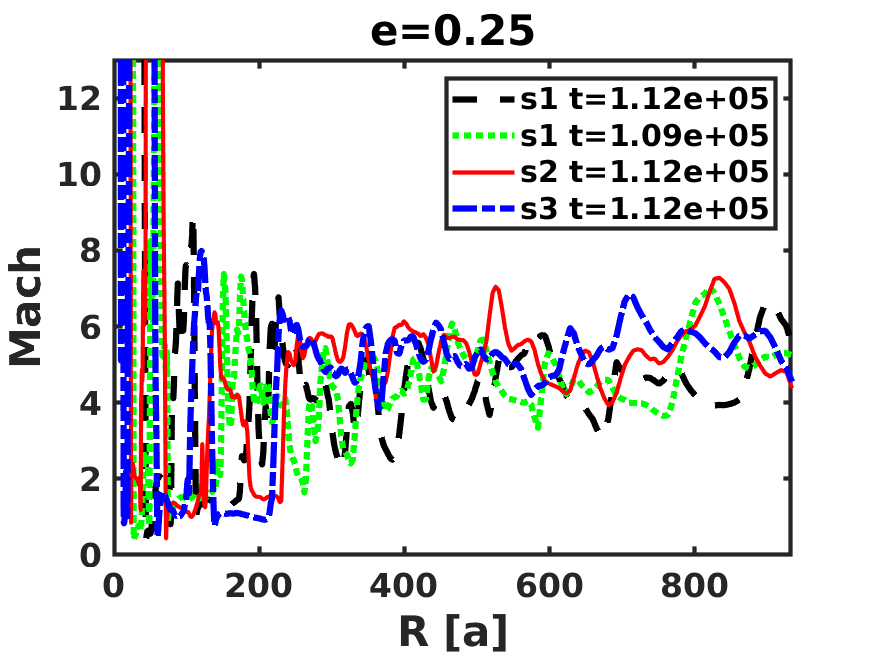}
\includegraphics[width=6.7 cm]{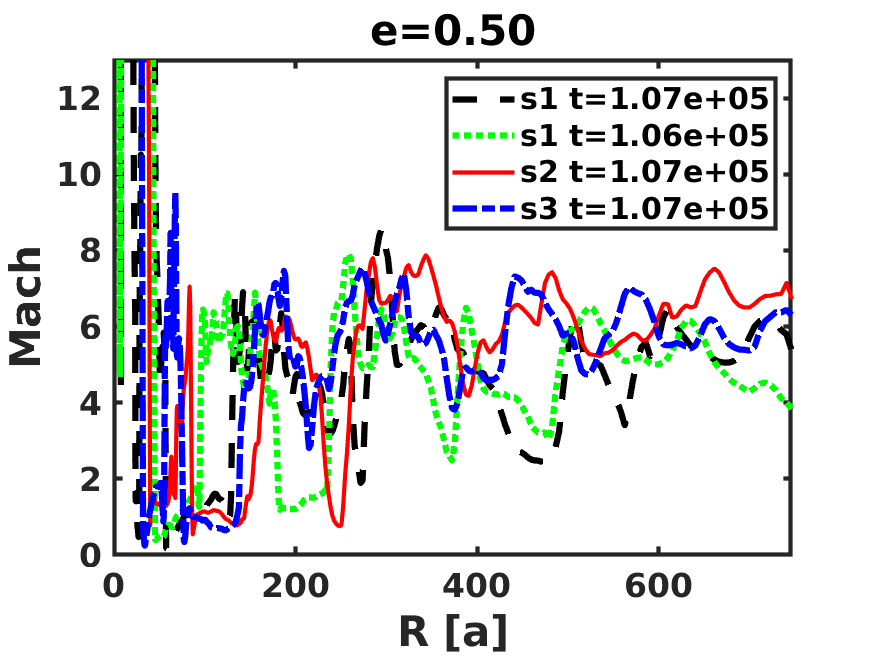}
\includegraphics[width=6.7 cm]{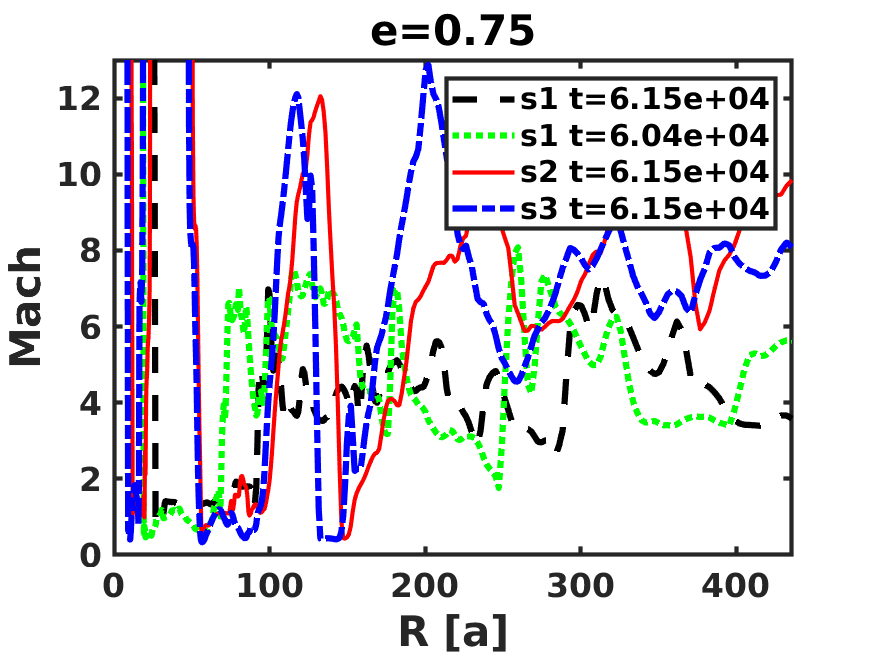}
\caption{The same as in Fig.~\ref{fig:vavR} but for Mach number.}
\label{fig:MachR}
\end{figure}

The radial distributions of $\varv_r$ averaged over s1, s2, and s3, at simulation time $T=1.11\times 10^5$ are presented in Fig.~\ref{fig:vavR}. We note that time in this work is given in simulation units, which are $(a/c)=200$~s and are implicit. The figure includes the case for s1 at $T=1.09\times 10^5$ as well to illustrate short-term variability on top of the longer term behavior. Independently of the eccentricity, there are strong spatial variations in the radial velocity up to $R\sim 300\,a$ that are related to the spiral interaction structure. Beyond $R\sim 300\,a$, there is a gradual acceleration of the flow, which is at that point already made of a mixture of pulsar and stellar winds. This acceleration is equally prominent for all sectors when $e=0$, although the higher the eccentricity, the stronger the acceleration becomes for s1, around the periastron-apastron direction, and weaker for s2 and s3. This is also seen in Fig.~\ref{fig:vave}, which displays the dependence of the sector-averaged $\varv_r$ with $e$ at $R=400\,a$. In general, the acceleration slows down at $R\sim 400-600\,a$ depending on $e$, which is expected as the flow becomes highly supersonic. In Fig.~\ref{fig:MachR}, the radial distribution of averaged Mach number is also shown. At smaller radii, $R\lesssim 300\,a$, large variations of this quantity are related to the spiral structure, whereas at larger radii the flow averaged Mach number approaches $\sim 6$ ({being still strongly variable for high $e$)}. Despite the shallower drop of pressure in s1 shown above (see Fig.~\ref{fig:vave}), $\varv_r$ grows faster in that direction because more energy is invested in the flow motion whereas the average density is lower.

\begin{figure}	
%\widefigure
\includegraphics[width=3.95 cm]{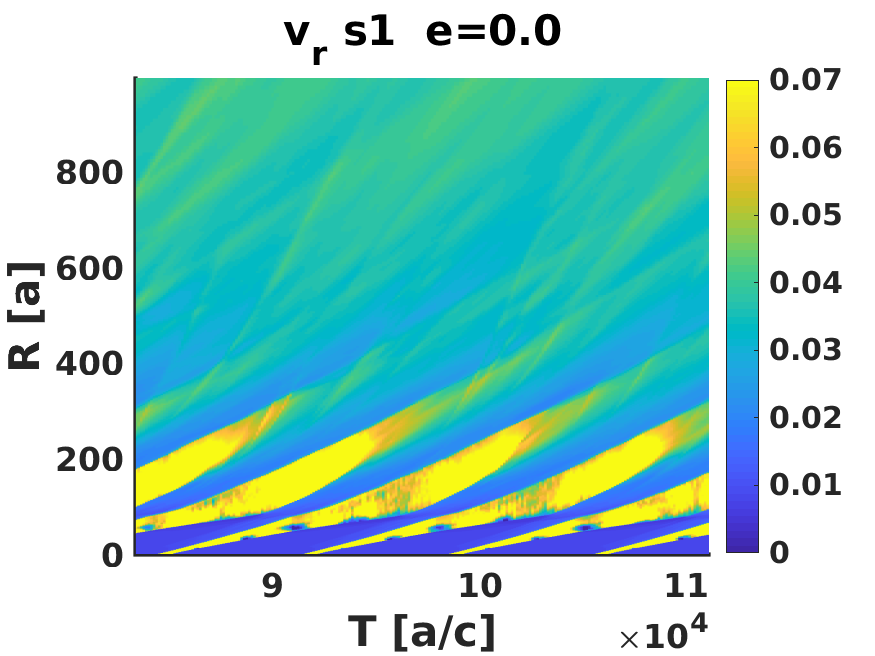}
\includegraphics[width=3.95 cm]{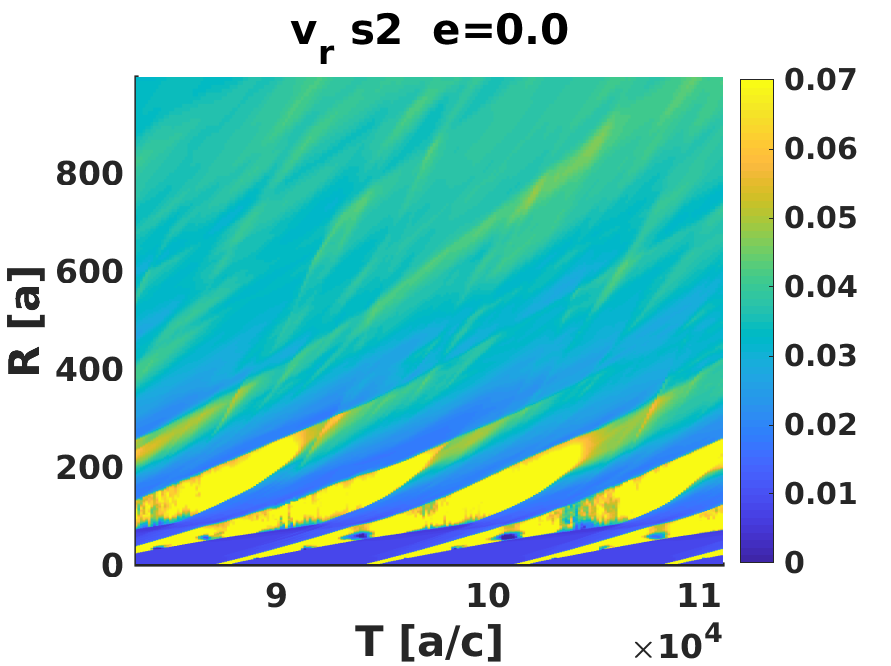}
\includegraphics[width=3.95 cm]{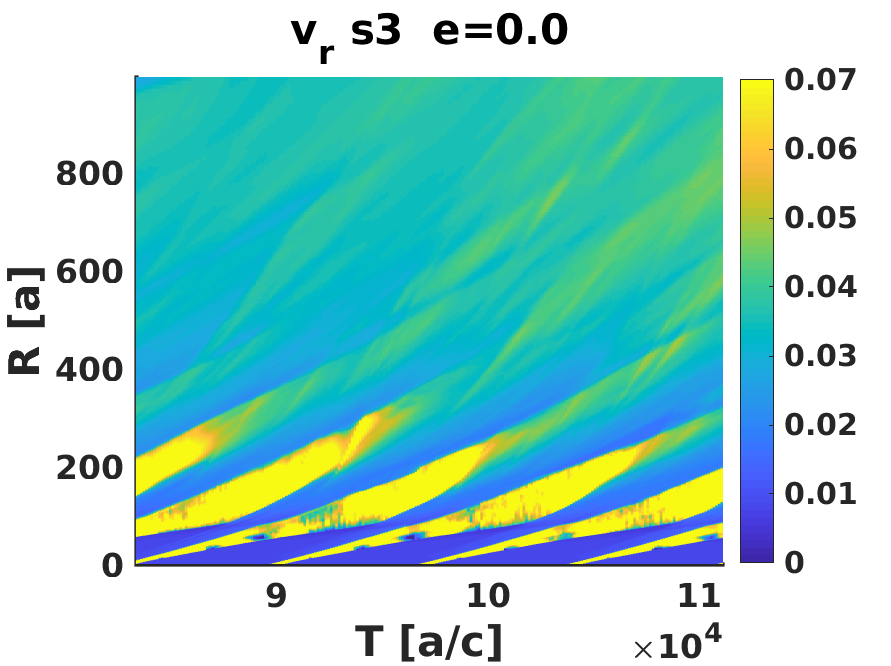}\\
\includegraphics[width=3.95 cm]{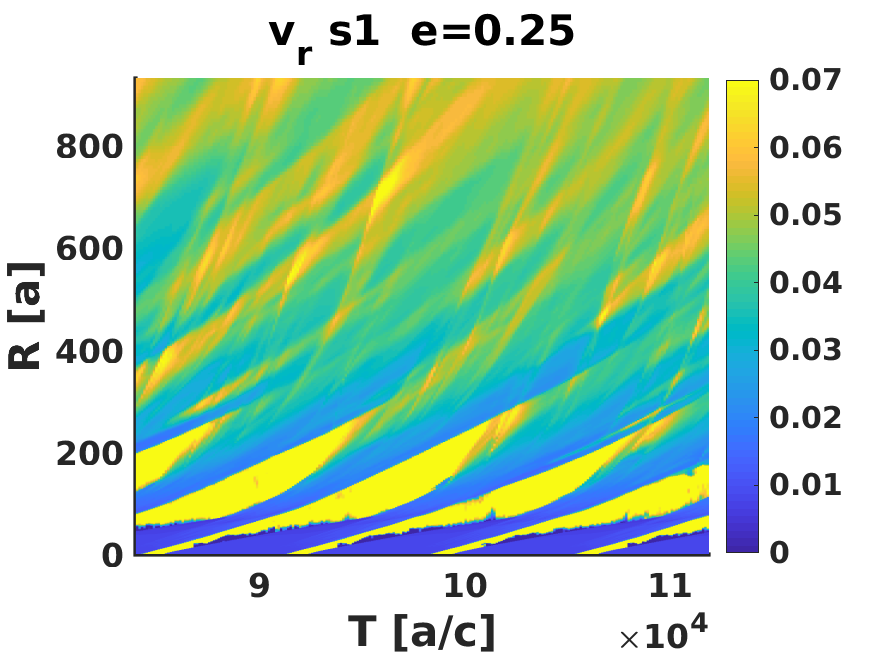}
\includegraphics[width=3.95 cm]{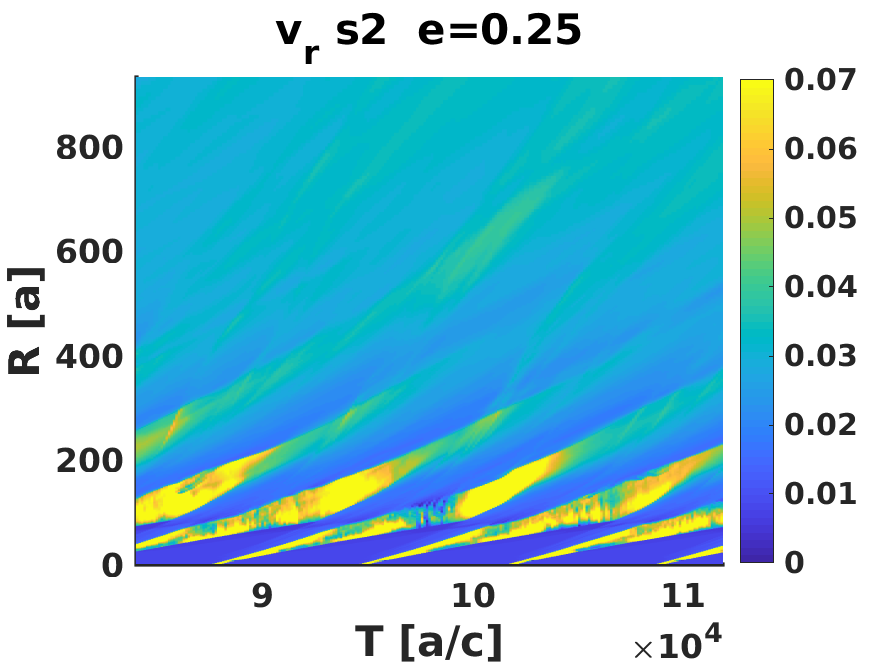}
\includegraphics[width=3.95 cm]{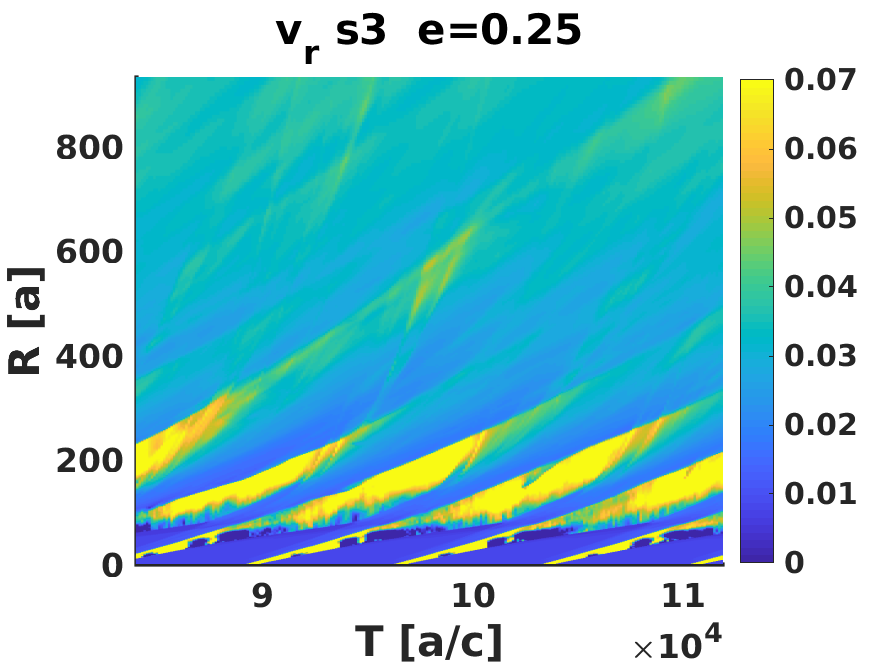}\\
\includegraphics[width=3.95 cm]{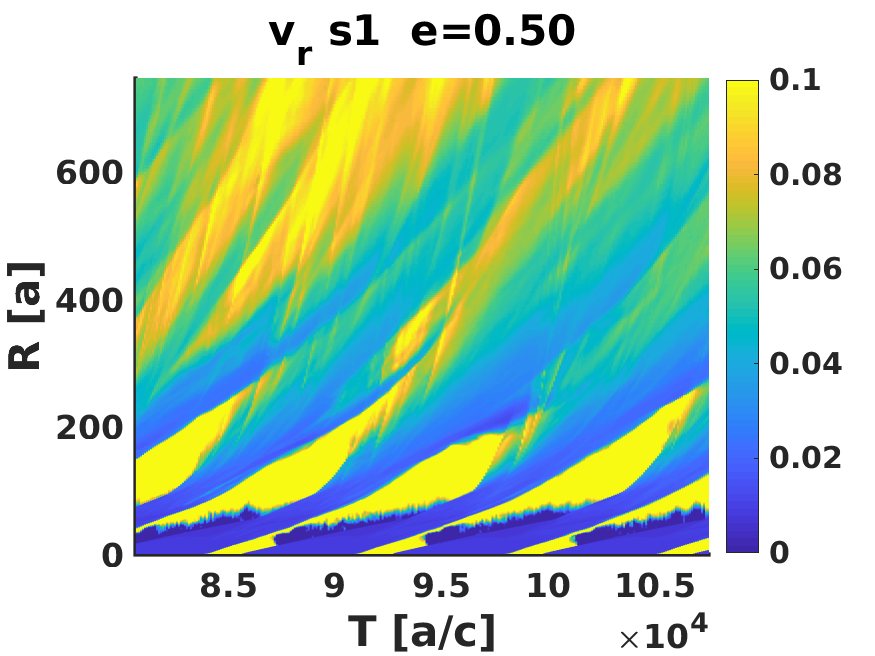}
\includegraphics[width=3.95 cm]{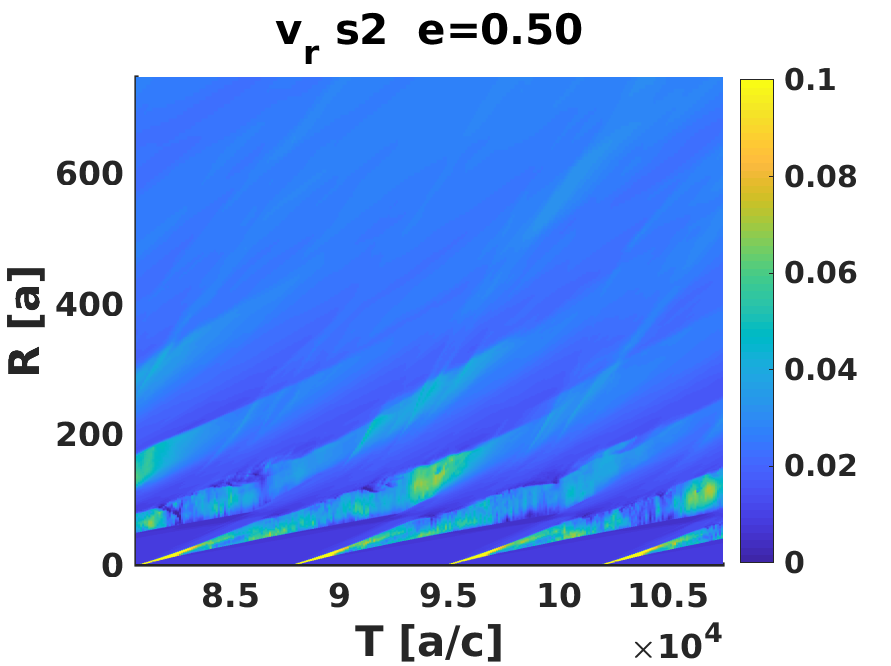}
\includegraphics[width=3.95 cm]{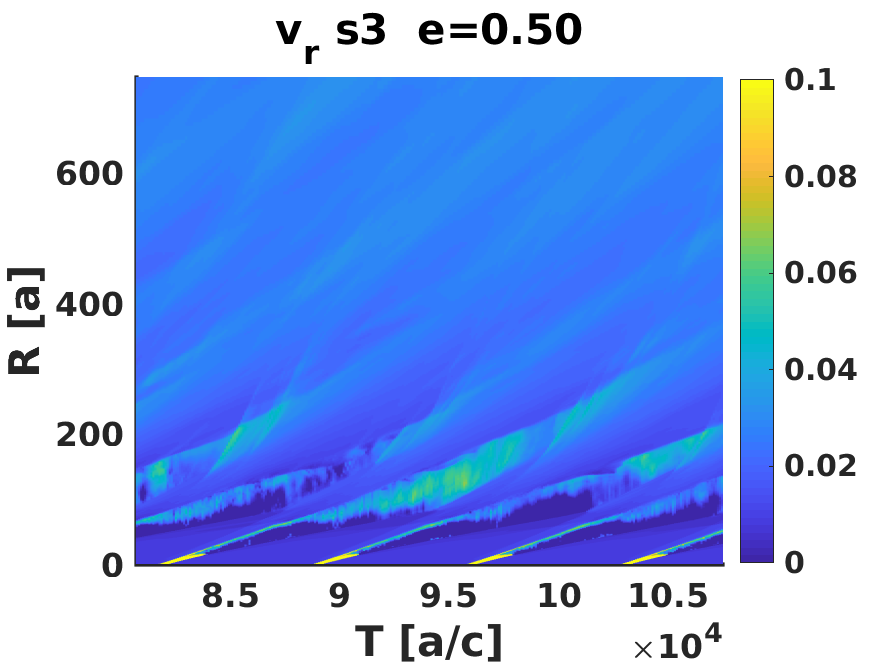}\\
\includegraphics[width=3.95 cm]{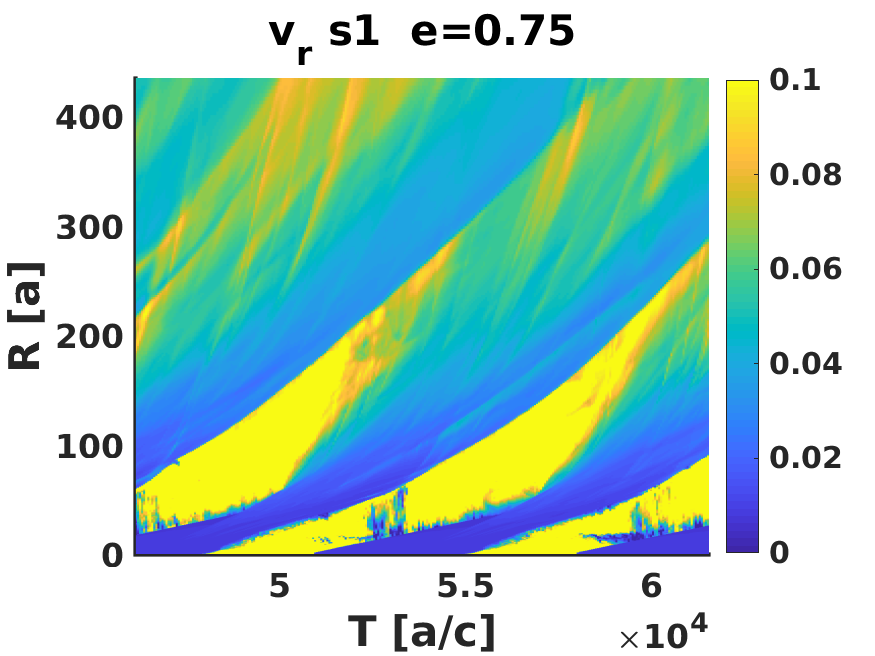}
\includegraphics[width=3.95 cm]{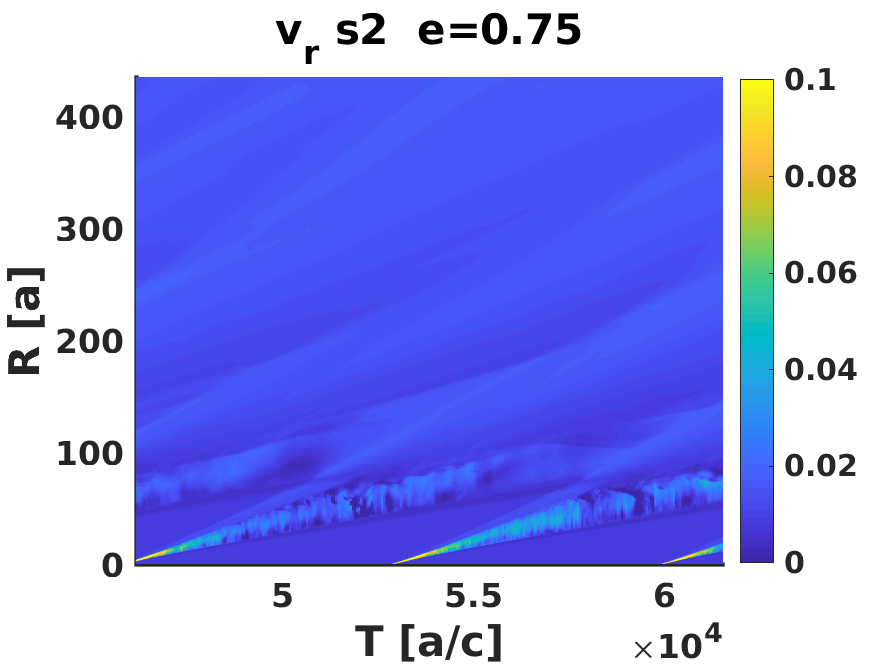}
\includegraphics[width=3.95 cm]{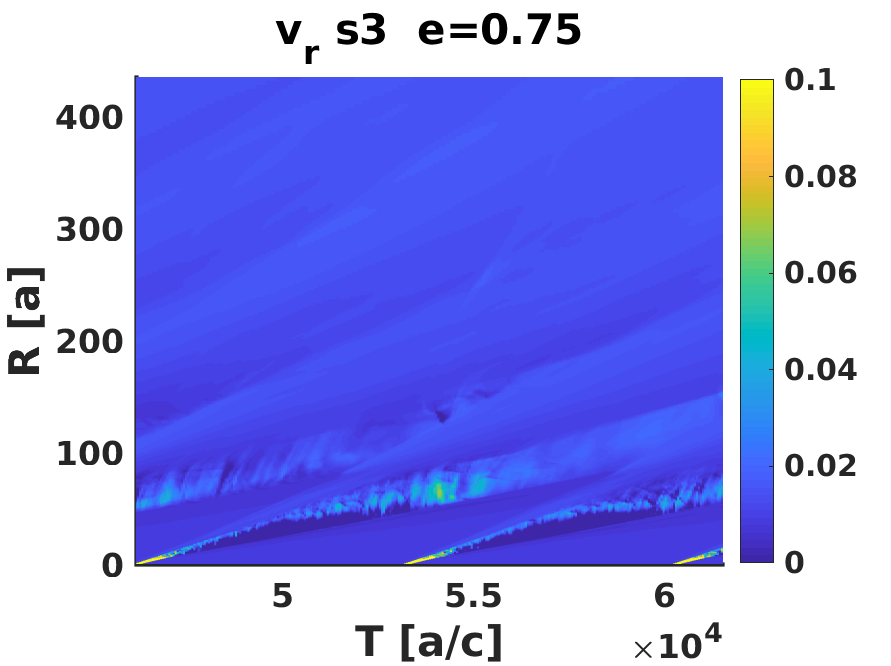}
\caption{Color map of $\varv_r$ {(in light speed units)}, averaged over sectors s1, s2 and s3, in the $R$ ($y$-axis) versus $T$ ($x$-axis) plane. Rows from top to bottom correspond to $e=0$, 0.25, 0.5 and 0.75, and columns from left to right to s1, s2 and s3.}
\label{fig:vavT}
\end{figure}

\begin{figure}	
\includegraphics[width=3.95 cm]{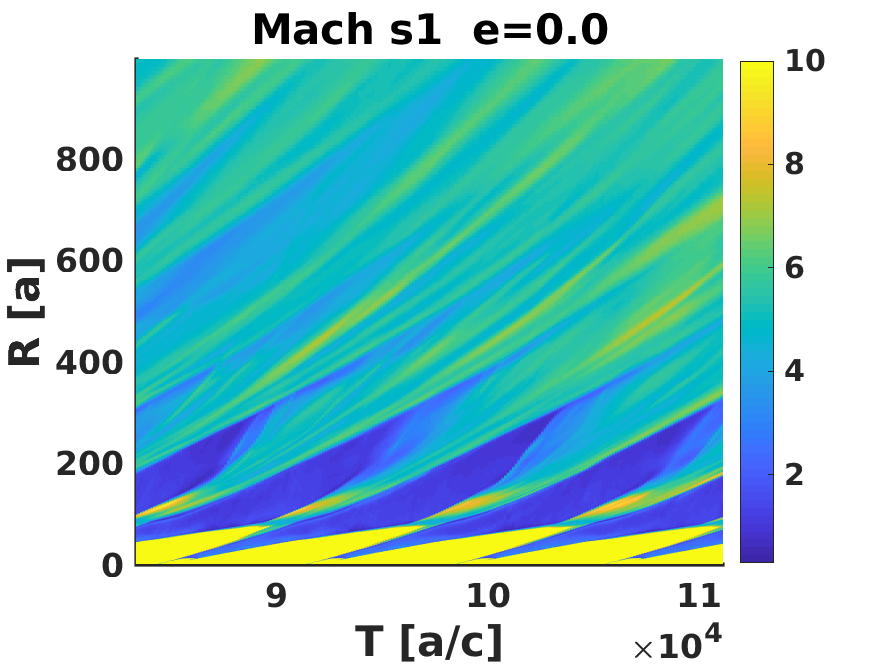}
\includegraphics[width=3.95 cm]{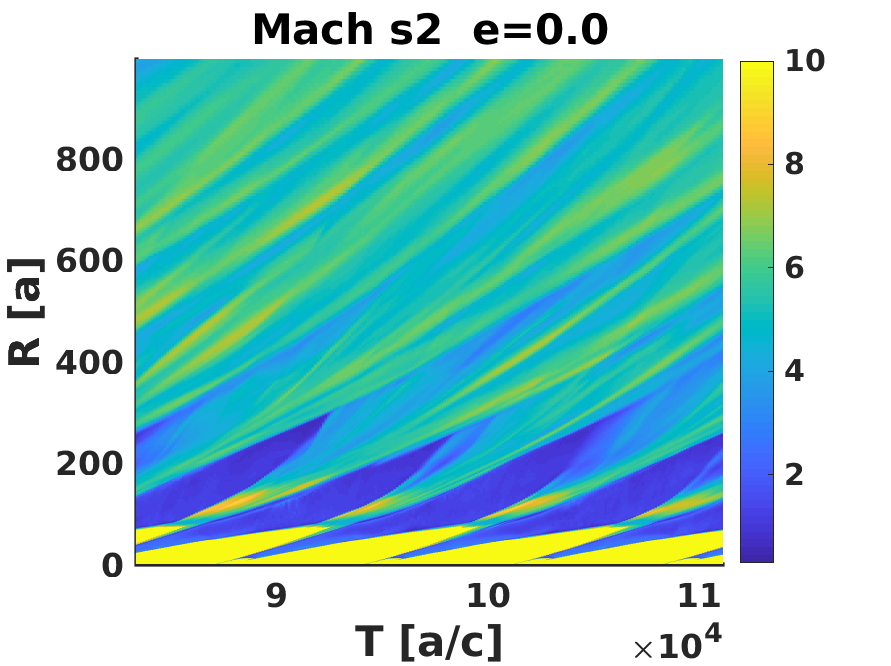}
\includegraphics[width=3.95 cm]{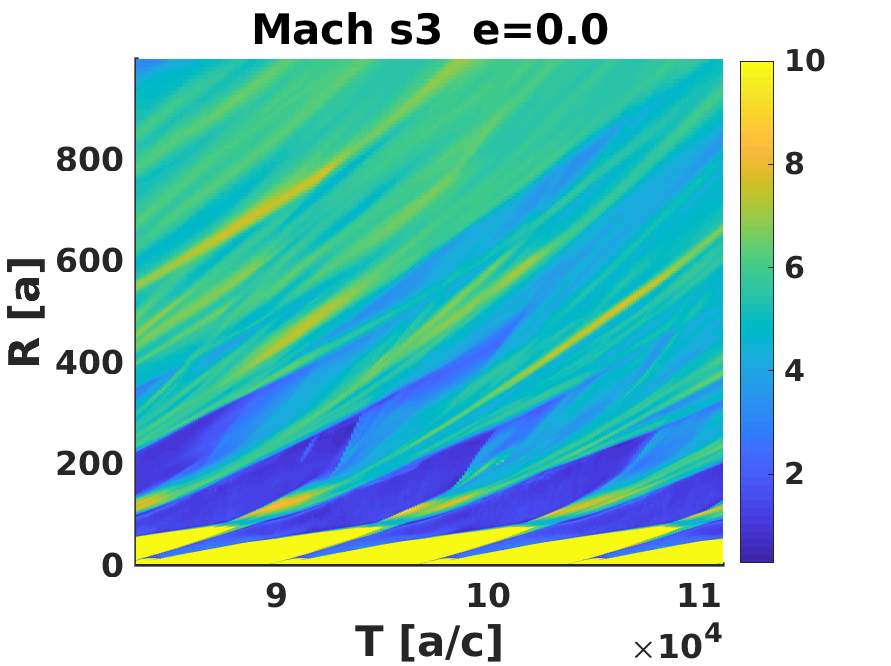}\\
\includegraphics[width=3.95 cm]{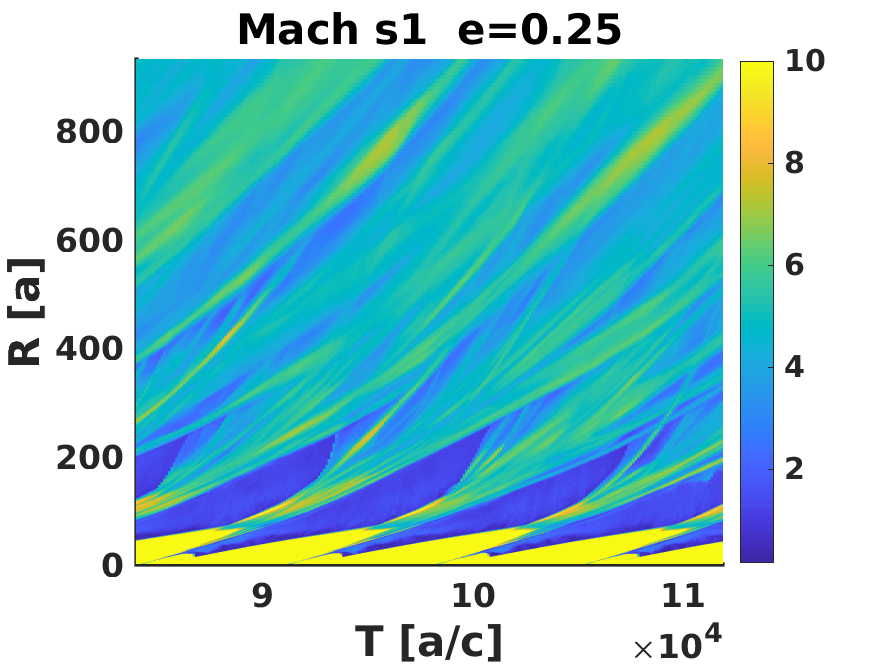}
\includegraphics[width=3.95 cm]{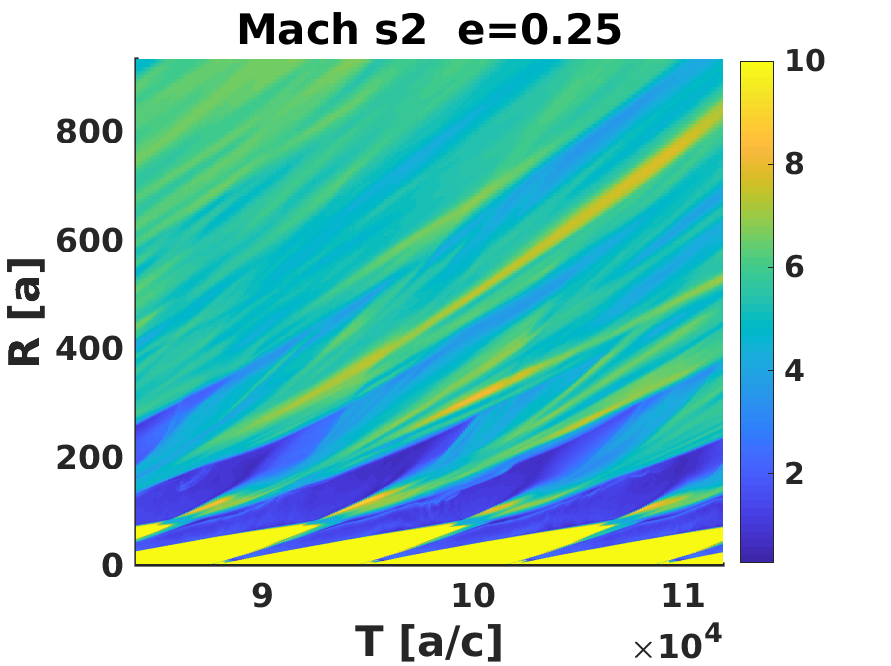}
\includegraphics[width=3.95 cm]{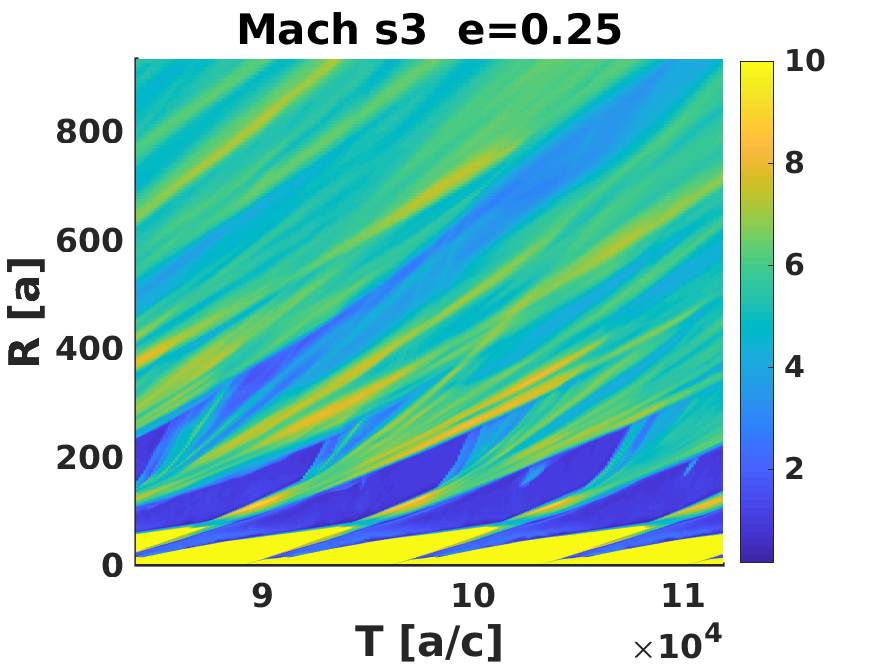}\\
\includegraphics[width=3.95 cm]{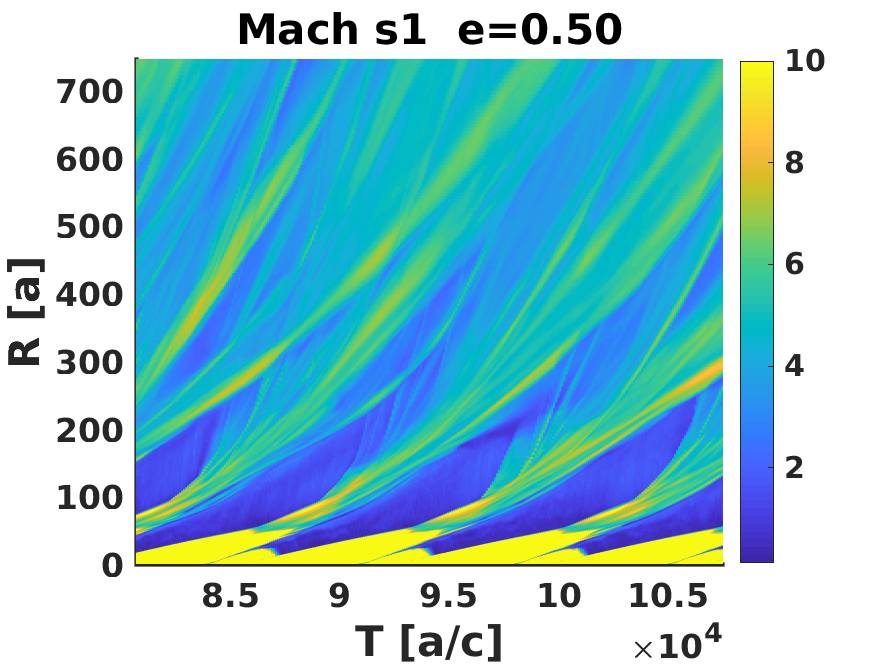}
\includegraphics[width=3.95 cm]{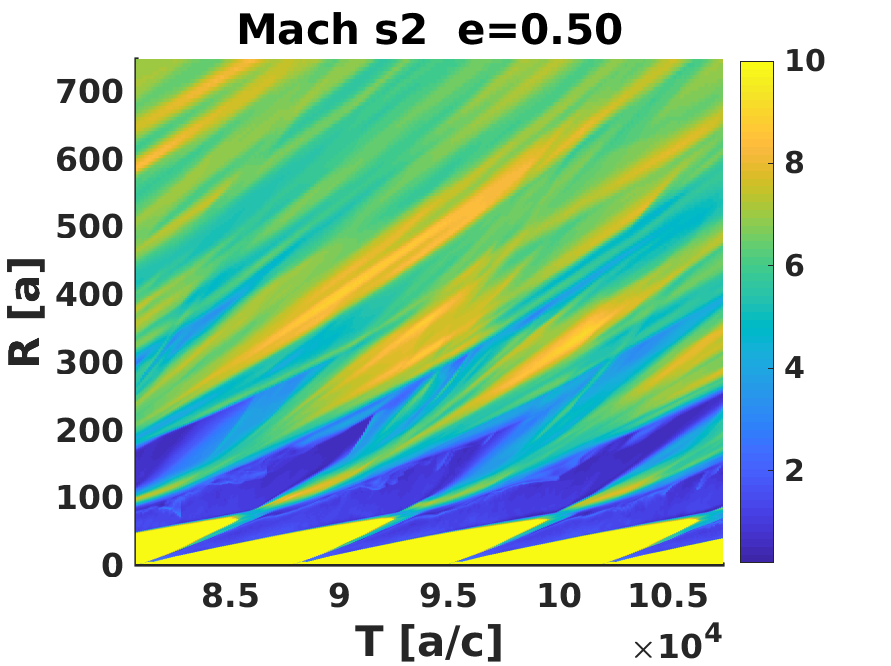}
\includegraphics[width=3.95 cm]{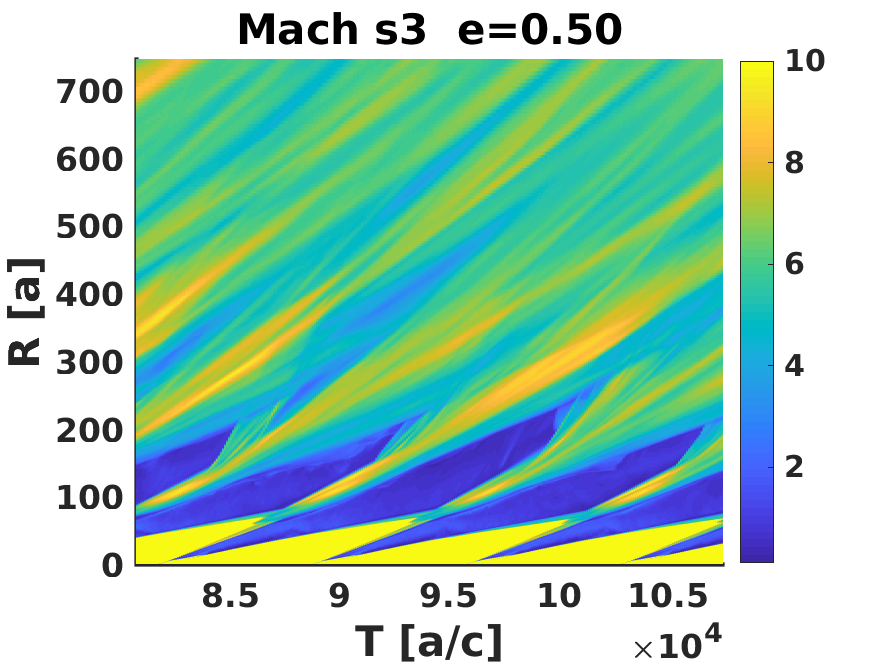}\\
\includegraphics[width=3.95 cm]{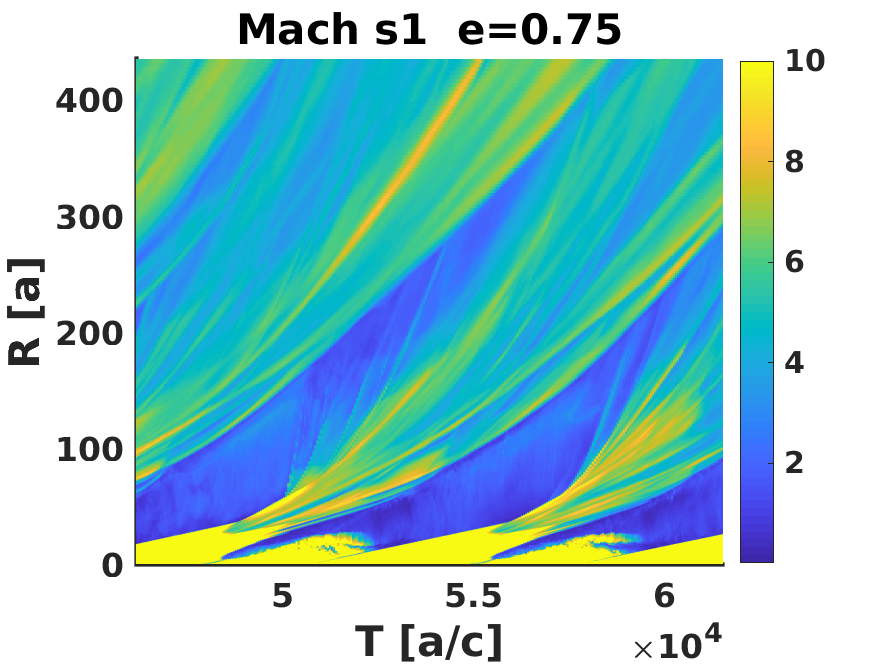}
\includegraphics[width=3.95 cm]{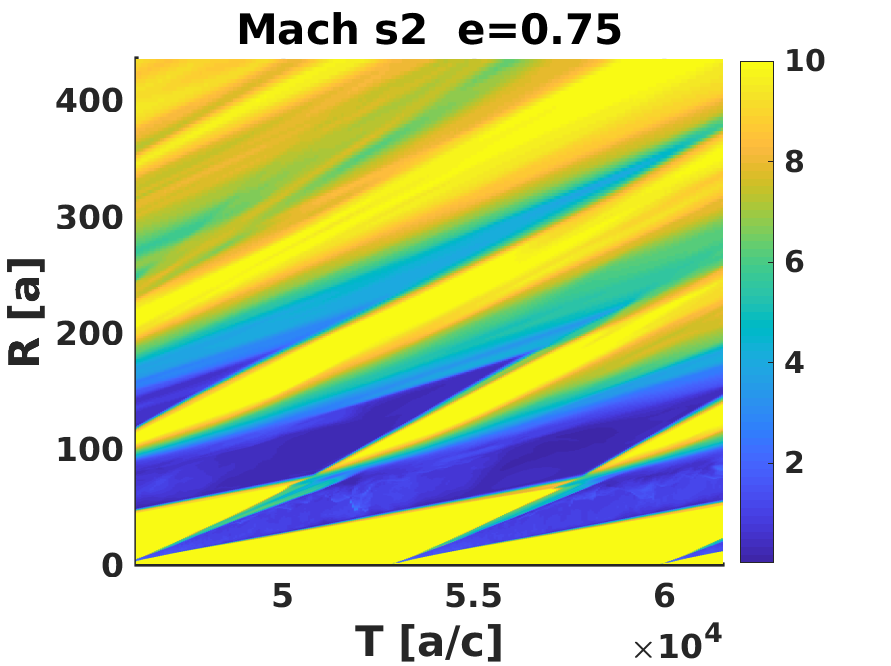}
\includegraphics[width=3.95 cm]{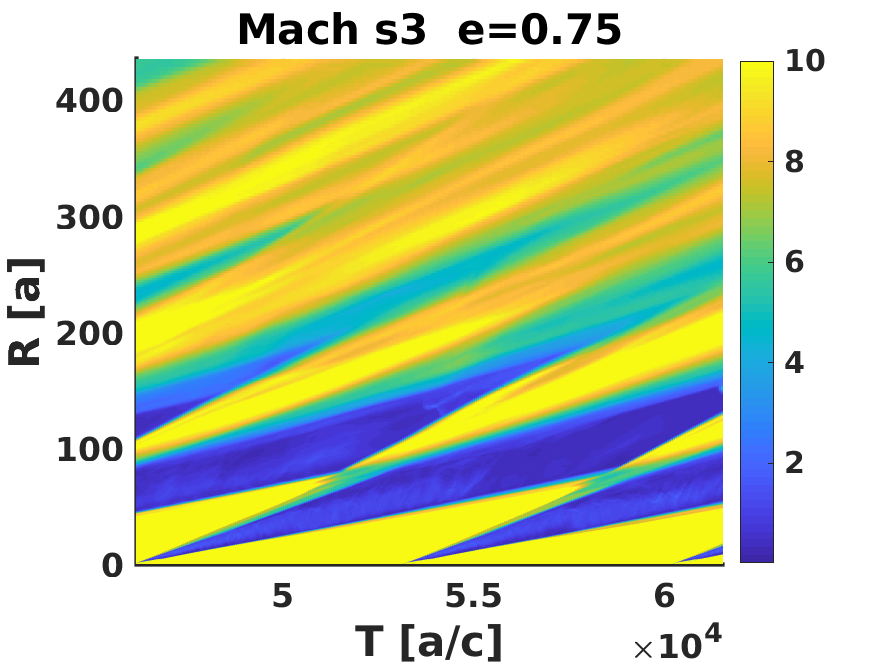}
\caption{The same as in Fig.~\ref{fig:vavT} but for Mach number.}
\label{fig:MachT}
\end{figure}

\begin{figure}	
%\widefigure
\includegraphics[width=6.7 cm]{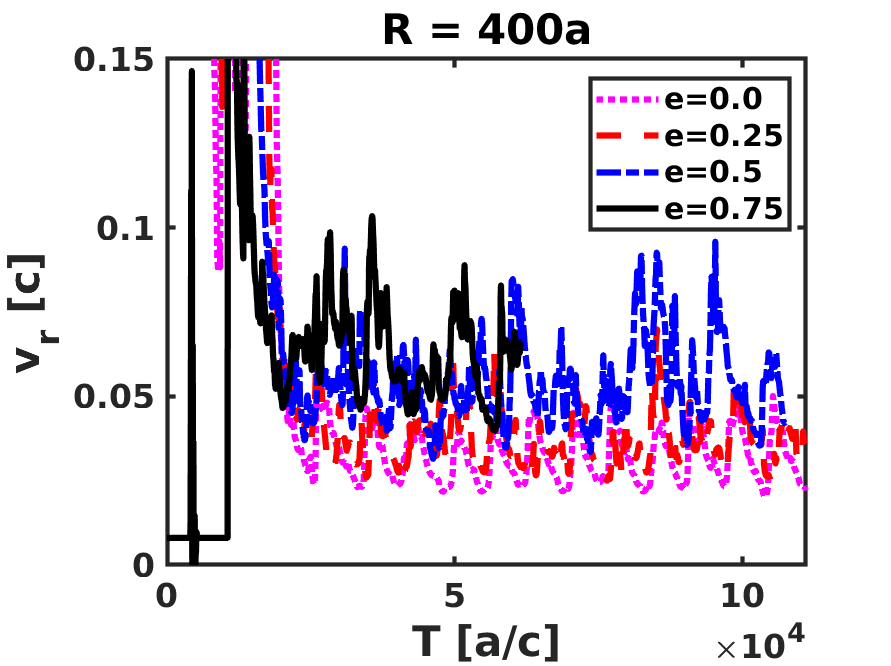}
\caption{Radial velocity, averaged over s1, versus $T$ at $R=400\,a$ for different eccentricities.}
\label{fig:vavTe}
\end{figure}

To illustrate the temporal evolution of the system, color maps of sector-averaged radial velocity and Mach number in the $R$ ($y$-axis) versus $T$ ($x$-axis) plane are presented in Figs.~\ref{fig:vavT} and \ref{fig:MachT}, respectively. These maps show two different colour regions that indicate the transition at $R<300\,a$ from a spiral structure to a more homogeneous, mixed, outflow at larger radii. This effect is very prominent in all directions for low eccentricities, both in averaged $\varv_r$ and Mach number, whereas for high eccentricities the same effect is much more prominent in s1 (around the periastron-apastron direction) than in s2 and s3 for $\varv_r$, whereas for Mach number the opposite happens, although less dramatically. The slower increase in the s1-averaged Mach number for high $e$-values is related to the presence of shocks that reheat the flow. The contrast of shocked flow behavior depending on $e$ is also illustrated in Fig.~\ref{fig:vavTe}, which shows the whole $\phi$-averaged in sector s1 $\varv_r$ versus $T$ at $R=400\,a$ for different eccentricities. The velocity jumps grows  with eccentricity and became more pronounced for $e\ge0.5$.

\begin{figure}	
%\widefigure
\includegraphics[width=6.7 cm]{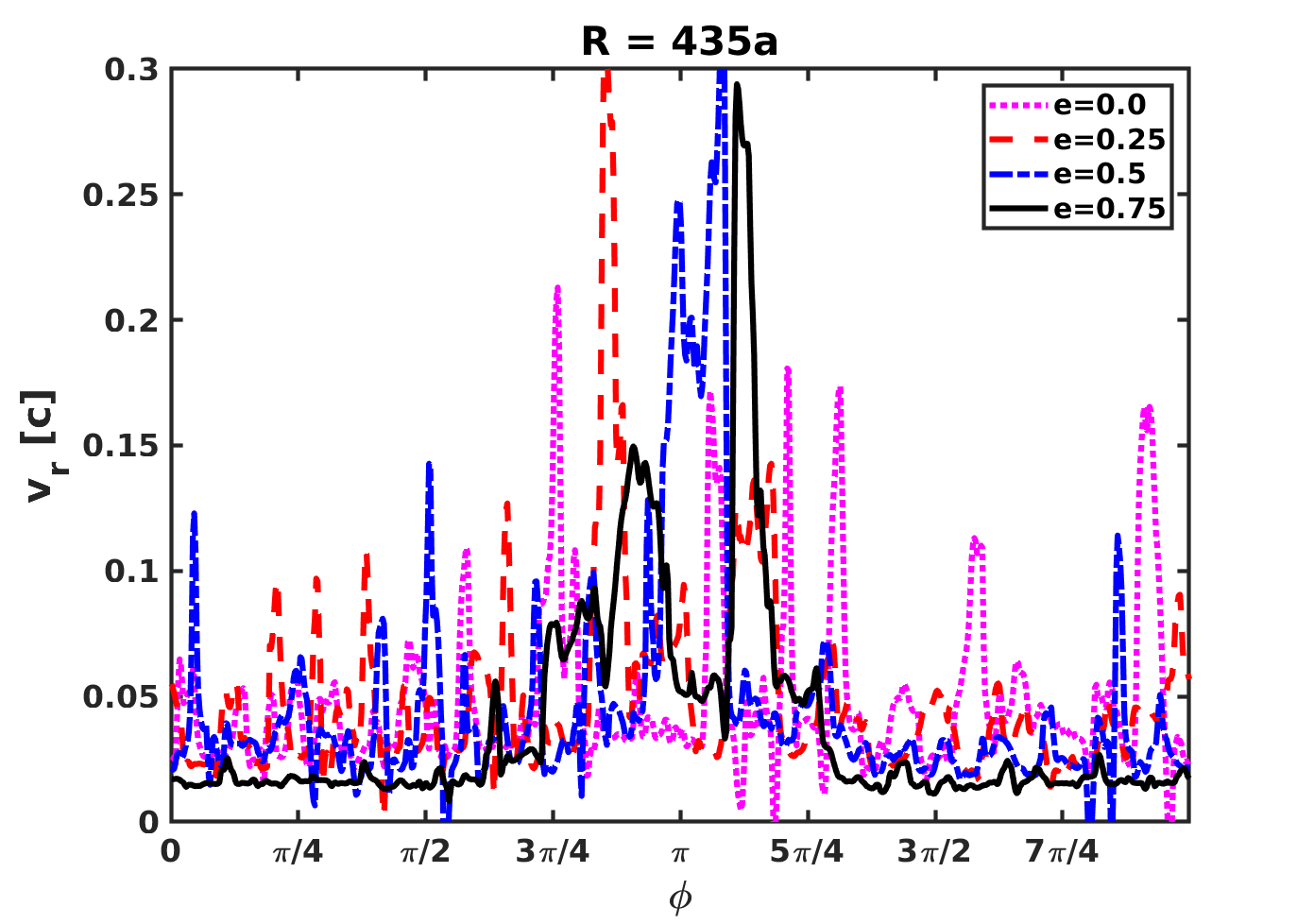}
\caption{Radial velocity, averaged over $R$ in the interval $[350\,a, 435\,a]$, versus $\phi$ for different eccentricities.}
\label{fig:vavphi}
\end{figure}

\begin{figure}	
%\widefigure
\includegraphics[width=6.7 cm]{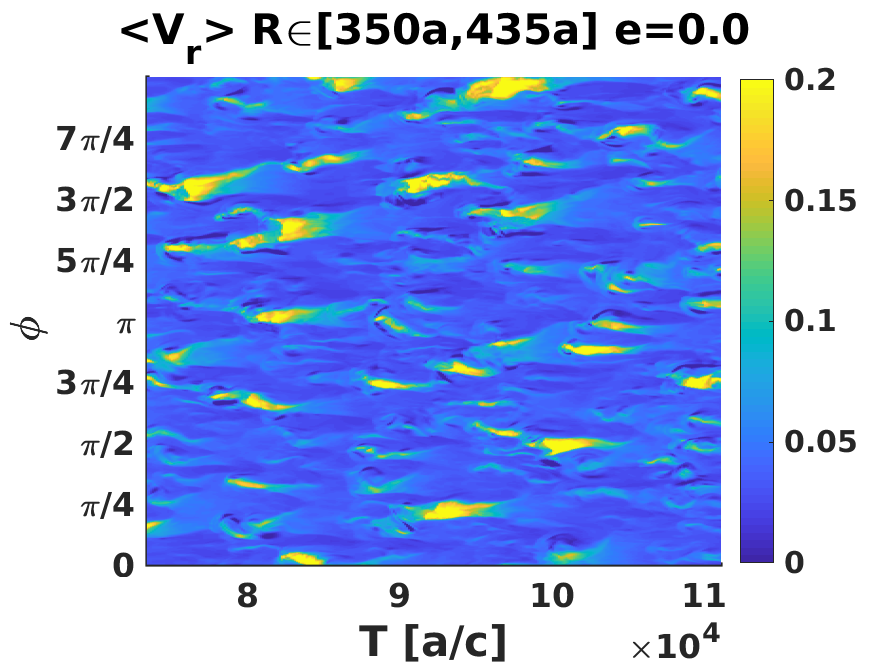}
\includegraphics[width=6.7 cm]{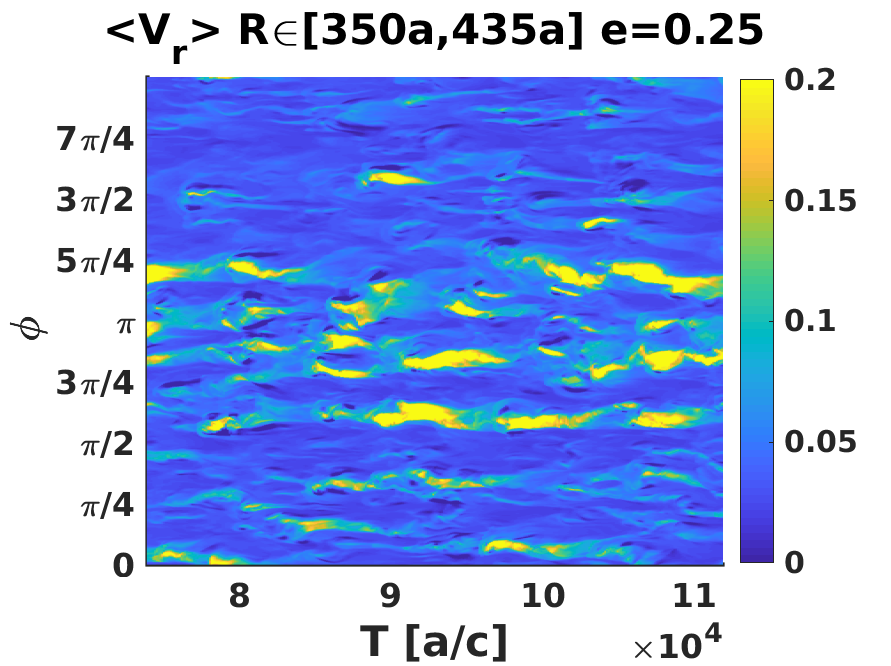}
\includegraphics[width=6.7 cm]{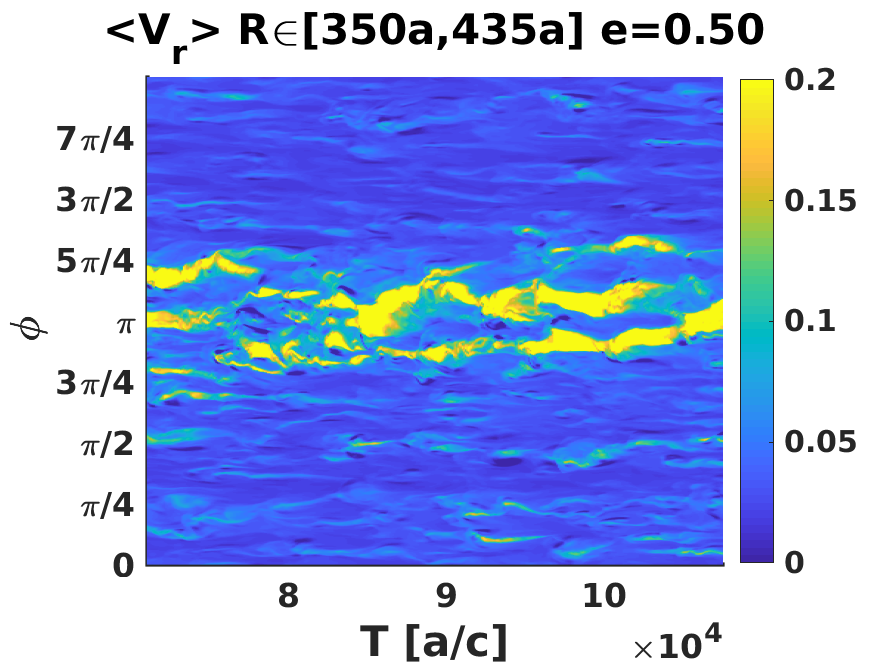}
\includegraphics[width=6.7 cm]{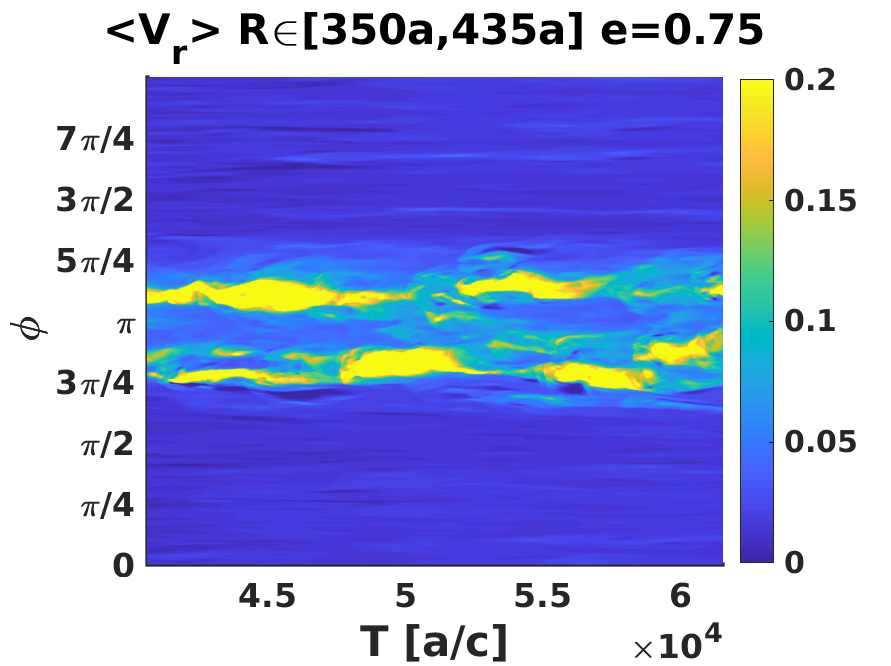}
\caption{Map of $\varv_r$, averaged over $R$ in the interval $[350\,a, 435\,a]$, in the $\phi$ ($y$-axis) versus $T$ ($x$-axis) plane, at $R=400\,a$, for different eccentricities: $e=0$ (left-top panel); $e=0.25$ (right-top panel); $e=0.5$ (left-bottom panel); and $e=0.75$ (right-bottom panel). }
\label{fig:vavphiT}
\end{figure}

The radial velocity, averaged over $R$ in the interval $[350\,a,435\,a]$, versus $\phi$ is shown in Fig.~\ref{fig:vavphi}. The radial velocity is also color mapped on the $\phi$ ($y$-axis) versus $T$ ($x$-axis) plane in Fig.~\ref{fig:vavphiT}. Consistently with previous figures, one sees that the low $e$ cases present a quasi-isotropic flow with significant stochastic behavior on top of the longer term behavior, whereas high $e$-cases show faster flows concentrate around the periastron-apastron direction ($\phi=\pi$). 

%%%%%%%%%%%%%%%%%%%%%%%%%%%%%%%%%%%%%%%%%%
\section{Summary and discussion}\label{dis}

In this article, we have presented a detailed analysis of the results obtained by the simulation of colliding pulsar and stellar winds. For the first time, cases with different eccentricities are systematically explored, so the impact of this prediction can be assessed. {The further development of these models should impact the studies of very powerful sources such as HMPB or microquasars \citep{2010MNRAS.404L..55D,2018MNRAS.479.4399Z,2021AN....342..337S,2020MNRAS.498.3592M}. We note that the first simulation of a jet-stellar wind interaction in a microquasar along a full orbit has been already done in \cite{brb21}.}

The evolution of the shocked pulsar and stellar winds was studied analytically in \cite{bb11}.
The mixed-wind eventual velocity away (expel) from the binary can be estimated as:
\be
\varv_{\rm exp} = \sqrt{\frac{2L_{sd}}{\dot{M}_w}} = \sqrt{2\eta \varv_w c} \approx 0.04 \;c \eta_{-1}^{1/2};
\label{eq:vexp}
\ee
{where $\eta_{-1} = \eta/10^{-1}$, and our numerical results, $\varv_{\rm exp}= 0.40\pm 0.03 \;c $ (see Fig.~\ref{fig:vavR}), confirm this prediction.}

In our numerical calculations, we find that the shocked flow structure on large scales is a slowly-accelerating, supersonic mixture of shocked stellar and pulsar winds, with an approximately isotropic propagation in $\phi$. We note that the simulations could not properly probe the expansion in $\theta$ due to low resolution, although 3D simulations by \citep{bbp15} suggest that on scales $\gg a$ the shocked structure can become wider in that direction than conical expansion predicts because of internal energy confinement. The present work also shows that this mixed supersonic wind is very clumpy in density and velocity, so particle acceleration may easily occur in such an environment. 

Our results are fully consistent with those obtained by \cite{bb16,bbmb17}, who showed for the first time the dramatic impact that eccentricity can have on the evolution of the shocked flows on large scales (non-thermal processes were discussed in \citealt{2018MNRAS.479.1320B}). In addition to that, the present work also characterizes the eccentricity of a transition between an approximately isotropic supersonic wind, made of shocked stellar and pulsar wind (for small $e$), and a sort of two-component structure, one fast, light and collimated, directed along the periastron-apastron direction, and the other slow, dense and broad, directed elsewhere (for high $e$). In particular, we find that the transition is somewhere between $e=0.5$ and 0.75, probably close to the latter. 

As mentioned in Sect.~\ref{model}, our results should not be sensitive to the orbital period on large scales. This implies that at a semi-quantitative level, our predictions can be extrapolated to wider systems. However, higher accuracy in the estimate of the
eccentricity associated to a structure geometry transition requires fully 3D calculations. On the other hand, the magnetic field could also play an important role in the evolution of the shocked flows, and should be included in future stages of this research. Moreover, future numerical work should tackle the issue of how the shocked mixed flows interact with the interstellar medium, both in the low $e$ \citep[see][for analytical predictions]{bos11} and the high $e$ regimes. Finally, the consequences of the eccentricity dependence of the shocked flow evolution with respect to non-thermal emission should be studied in more detail than what has been done so far.

%%%%%%%%%%%%%%%%%%%%%%%%%%%%%%%%%%%%%%%%%%
\vspace{6pt}

%%%%%%%%%%%%%%%%%%%%%%%%%%%%%%%%%%%%%%%%%%
{All authors have read and agreed to the published version of the manuscript. Barkov M. performed the numerical simulation and data analysis. Bosch-Ramon and Barkov worked on the text of the manuscript.}

\vspace{6pt} 

\subsection*{Foundlings}
{V.B-R. acknowledges financial support by the State Agency for Research of the Spanish Ministry of Science and Innovation under grant PID2019-105510GB-C31 and through the ''Unit of Excellence Mar\'ia de Maeztu 2020-2023'' award to the Institute of Cosmos Sciences (CEX2019-000918-M), and by the Catalan DEC grant 2017 SGR 643. }

\vspace{6pt} 

\subsection*{Data availability}
{The original data and its analysis can be requested by email.} 

\vspace{6pt} 

\subsection*{Acknowledgments}
{V.B-R. is Correspondent Researcher of CONICET, Argentina, at the IAR. The simulations were performed on the CFCA XC30 cluster of the National Astronomical Observatory of Japan.}

%\section*{References}

%\externalbibliography{yes}
%\bibliography{text}

\begin{thebibliography}{999}
	
	\bibitem[{Dubus}(2013)]{dub13}
	{Dubus}, G.
	\newblock {Gamma-ray binaries and related systems}.
	\newblock {\em A\&A~Rev} {\bf 2013}, {\em 21},~64,
	\href{http://xxx.lanl.gov/abs/1307.7083}{{\normalfont
			[arXiv:astro-ph.HE/1307.7083]}}.
	\newblock
	doi:{\changeurlcolor{black}\href{https://doi.org/10.1007/s00159-013-0064-5}{\detokenize{10.1007/s00159-013-0064-5}}}.
	
	\bibitem[{Paredes} and {Bordas}(2019{\natexlab{a}})]{par19a}
	{Paredes}, J.M.; {Bordas}, P.
	\newblock {Phenomenology of gamma-ray emitting binaries}.
	\newblock {\em arXiv e-prints} {\bf 2019}, p. arXiv:1902.09898,
	\href{http://xxx.lanl.gov/abs/1902.09898}{{\normalfont
			[arXiv:astro-ph.HE/1902.09898]}}.
	
	\bibitem[{Paredes} and {Bordas}(2019{\natexlab{b}})]{par19b}
	{Paredes}, J.M.; {Bordas}, P.
	\newblock {Broad-band Emission from Gamma-ray Binaries}.
	\newblock  Frontier Research in Astrophysics - III. 28 May - 2 June 2018.
	Mondello (Palermo,  2019, p.~44,
	\href{http://xxx.lanl.gov/abs/1901.03624}{{\normalfont
			[arXiv:astro-ph.HE/1901.03624]}}.
	
	\bibitem[{Tavani} and {Arons}(1997)]{tav97}
	{Tavani}, M.; {Arons}, J.
	\newblock {Theory of High-Energy Emission from the Pulsar/Be Star System PSR
		1259-63. I. Radiation Mechanisms and Interaction Geometry}.
	\newblock {\em \apj} {\bf 1997}, {\em 477},~439--464,
	\href{http://xxx.lanl.gov/abs/astro-ph/9609086}{{\normalfont
			[arXiv:astro-ph/astro-ph/9609086]}}.
	\newblock
	doi:{\changeurlcolor{black}\href{https://doi.org/10.1086/303676}{\detokenize{10.1086/303676}}}.
	
	\bibitem[{Sierpowska} and {Bednarek}(2005)]{sie05}
	{Sierpowska}, A.; {Bednarek}, W.
	\newblock {{\ensuremath{\gamma}}-rays from cascades in close massive binaries
		containing energetic pulsars}.
	\newblock {\em \mnras} {\bf 2005}, {\em 356},~711--726,
	\href{http://xxx.lanl.gov/abs/astro-ph/0410304}{{\normalfont
			[arXiv:astro-ph/astro-ph/0410304]}}.
	\newblock
	doi:{\changeurlcolor{black}\href{https://doi.org/10.1111/j.1365-2966.2004.08490.x}{\detokenize{10.1111/j.1365-2966.2004.08490.x}}}.
	
	\bibitem[{Dubus}(2006)]{dub06}
	{Dubus}, G.
	\newblock {Gamma-ray binaries: pulsars in disguise?}
	\newblock {\em \aap} {\bf 2006}, {\em 456},~801--817,
	\href{http://xxx.lanl.gov/abs/astro-ph/0605287}{{\normalfont
			[arXiv:astro-ph/astro-ph/0605287]}}.
	\newblock
	doi:{\changeurlcolor{black}\href{https://doi.org/10.1051/0004-6361:20054779}{\detokenize{10.1051/0004-6361:20054779}}}.
	
	\bibitem[{Neronov} and {Chernyakova}(2007)]{ner07}
	{Neronov}, A.; {Chernyakova}, M.
	\newblock {Radio-to-TeV {$\gamma$}-ray emission from PSR B1259 63}.
	\newblock {\em \apss} {\bf 2007}, {\em 309},~253--259,
	\href{http://xxx.lanl.gov/abs/astro-ph/0610139}{{\normalfont
			[astro-ph/0610139]}}.
	\newblock
	doi:{\changeurlcolor{black}\href{https://doi.org/10.1007/s10509-007-9454-3}{\detokenize{10.1007/s10509-007-9454-3}}}.
	
	\bibitem[{Khangulyan} \em{et~al.}(2007){Khangulyan}, {Hnatic}, {Aharonian}, and
	{Bogovalov}]{kha07}
	{Khangulyan}, D.; {Hnatic}, S.; {Aharonian}, F.; {Bogovalov}, S.
	\newblock {TeV light curve of PSR B1259-63/SS2883}.
	\newblock {\em \mnras} {\bf 2007}, {\em 380},~320--330,
	\href{http://xxx.lanl.gov/abs/arXiv:astro-ph/0605663}{{\normalfont
			[arXiv:astro-ph/0605663]}}.
	\newblock
	doi:{\changeurlcolor{black}\href{https://doi.org/10.1111/j.1365-2966.2007.12075.x}{\detokenize{10.1111/j.1365-2966.2007.12075.x}}}.
	
	\bibitem[{Kong} \em{et~al.}(2012){Kong}, {Cheng}, and {Huang}]{kon12}
	{Kong}, S.W.; {Cheng}, K.S.; {Huang}, Y.F.
	\newblock {Modeling the Multiwavelength Light Curves of PSR B1259-63/LS 2883.
		II. The Effects of Anisotropic Pulsar Wind and Doppler Boosting}.
	\newblock {\em \apj} {\bf 2012}, {\em 753},~127,
	\href{http://xxx.lanl.gov/abs/1205.2147}{{\normalfont
			[arXiv:astro-ph.HE/1205.2147]}}.
	\newblock
	doi:{\changeurlcolor{black}\href{https://doi.org/10.1088/0004-637X/753/2/127}{\detokenize{10.1088/0004-637X/753/2/127}}}.
	
	\bibitem[{Zabalza} \em{et~al.}(2013){Zabalza}, {Bosch-Ramon}, {Aharonian}, and
	{Khangulyan}]{zab13}
	{Zabalza}, V.; {Bosch-Ramon}, V.; {Aharonian}, F.; {Khangulyan}, D.
	\newblock {Unraveling the high-energy emission components of gamma-ray
		binaries}.
	\newblock {\em \aap} {\bf 2013}, {\em 551},~A17,
	\href{http://xxx.lanl.gov/abs/1212.3222}{{\normalfont
			[arXiv:astro-ph.HE/1212.3222]}}.
	\newblock
	doi:{\changeurlcolor{black}\href{https://doi.org/10.1051/0004-6361/201220589}{\detokenize{10.1051/0004-6361/201220589}}}.
	
	\bibitem[{Dubus} \em{et~al.}(2015){Dubus}, {Lamberts}, and {Fromang}]{dub15}
	{Dubus}, G.; {Lamberts}, A.; {Fromang}, S.
	\newblock {Modelling the high-energy emission from gamma-ray binaries using
		numerical relativistic hydrodynamics}.
	\newblock {\em \aap} {\bf 2015}, {\em 581},~A27,
	\href{http://xxx.lanl.gov/abs/1505.01026}{{\normalfont
			[arXiv:astro-ph.HE/1505.01026]}}.
	\newblock
	doi:{\changeurlcolor{black}\href{https://doi.org/10.1051/0004-6361/201425394}{\detokenize{10.1051/0004-6361/201425394}}}.
	
	\bibitem[{Molina} and {Bosch-Ramon}(2020)]{mol20}
	{Molina}, E.; {Bosch-Ramon}, V.
	\newblock {A dynamical and radiation semi-analytical model of pulsar-star
		colliding winds along the orbit: Application to LS 5039}.
	\newblock {\em \aap} {\bf 2020}, {\em 641},~A84,
	\href{http://xxx.lanl.gov/abs/2007.00543}{{\normalfont
			[arXiv:astro-ph.HE/2007.00543]}}.
	\newblock
	doi:{\changeurlcolor{black}\href{https://doi.org/10.1051/0004-6361/202038417}{\detokenize{10.1051/0004-6361/202038417}}}.
	
	\bibitem[{Huber} \em{et~al.}(2021){Huber}, {Kissmann}, and {Reimer}]{hub21b}
	{Huber}, D.; {Kissmann}, R.; {Reimer}, O.
	\newblock {Relativistic fluid modelling of gamma-ray binaries. II. Application
		to LS 5039}.
	\newblock {\em \aap} {\bf 2021}, {\em 649},~A71,
	\href{http://xxx.lanl.gov/abs/2103.00995}{{\normalfont
			[arXiv:astro-ph.HE/2103.00995]}}.
	\newblock
	doi:{\changeurlcolor{black}\href{https://doi.org/10.1051/0004-6361/202039278}{\detokenize{10.1051/0004-6361/202039278}}}.
	
	\bibitem[{Lyutikov} \em{et~al.}(2020){Lyutikov}, {Barkov}, and
	{Giannios}]{2020ApJ...893L..39L}
	{Lyutikov}, M.; {Barkov}, M.V.; {Giannios}, D.
	\newblock {FRB Periodicity: Mild Pulsars in Tight O/B-star Binaries}.
	\newblock {\em \apjl} {\bf 2020}, {\em 893},~L39,
	\href{http://xxx.lanl.gov/abs/2002.01920}{{\normalfont
			[arXiv:astro-ph.HE/2002.01920]}}.
	\newblock
	doi:{\changeurlcolor{black}\href{https://doi.org/10.3847/2041-8213/ab87a4}{\detokenize{10.3847/2041-8213/ab87a4}}}.
	
	\bibitem[{Khangulyan} \em{et~al.}(2021){Khangulyan}, {Barkov}, and
	{Popov}]{2021arXiv210609858K}
	{Khangulyan}, D.; {Barkov}, M.V.; {Popov}, S.B.
	\newblock {High-frequency radio synchrotron maser emission from relativistic
		shocks}.
	\newblock {\em arXiv e-prints} {\bf 2021}, p. arXiv:2106.09858,
	\href{http://xxx.lanl.gov/abs/2106.09858}{{\normalfont
			[arXiv:astro-ph.HE/2106.09858]}}.
	
	\bibitem[{Bogovalov} \em{et~al.}(2008){Bogovalov}, {Khangulyan}, {Koldoba},
	{Ustyugova}, and {Aharonian}]{bog08}
	{Bogovalov}, S.V.; {Khangulyan}, D.V.; {Koldoba}, A.V.; {Ustyugova}, G.V.;
	{Aharonian}, F.A.
	\newblock {Modelling interaction of relativistic and non-relativistic winds in
		binary system PSR B1259-63/SS2883 - I. Hydrodynamical limit}.
	\newblock {\em \mnras} {\bf 2008}, {\em 387},~63--72,
	\href{http://xxx.lanl.gov/abs/0710.1961}{{\normalfont [0710.1961]}}.
	\newblock
	doi:{\changeurlcolor{black}\href{https://doi.org/10.1111/j.1365-2966.2008.13226.x}{\detokenize{10.1111/j.1365-2966.2008.13226.x}}}.
	
	\bibitem[{Eichler} and {Usov}(1993)]{1993ApJ...402..271E}
	{Eichler}, D.; {Usov}, V.
	\newblock {Particle Acceleration and Nonthermal Radio Emission in Binaries of
		Early-Type Stars}.
	\newblock {\em \apj} {\bf 1993}, {\em 402},~271.
	\newblock
	doi:{\changeurlcolor{black}\href{https://doi.org/10.1086/172130}{\detokenize{10.1086/172130}}}.
	
	\bibitem[{Bosch-Ramon} \em{et~al.}(2015){Bosch-Ramon}, {Barkov}, and
	{Perucho}]{bbp15}
	{Bosch-Ramon}, V.; {Barkov}, M.V.; {Perucho}, M.
	\newblock {Orbital evolution of colliding star and pulsar winds in 2D and 3D:
		effects of dimensionality, EoS, resolution, and grid size}.
	\newblock {\em \aap} {\bf 2015}, {\em 577},~A89,
	\href{http://xxx.lanl.gov/abs/1411.7892}{{\normalfont
			[arXiv:astro-ph.HE/1411.7892]}}.
	\newblock
	doi:{\changeurlcolor{black}\href{https://doi.org/10.1051/0004-6361/201425228}{\detokenize{10.1051/0004-6361/201425228}}}.
	
	\bibitem[{Bogovalov} \em{et~al.}(2012){Bogovalov}, {Khangulyan}, {Koldoba},
	{Ustyugova}, and {Aharonian}]{bog12}
	{Bogovalov}, S.V.; {Khangulyan}, D.; {Koldoba}, A.V.; {Ustyugova}, G.V.;
	{Aharonian}, F.A.
	\newblock {Modelling the interaction between relativistic and non-relativistic
		winds in the binary system PSR B1259-63/SS2883- II. Impact of the
		magnetization and anisotropy of the pulsar wind}.
	\newblock {\em \mnras} {\bf 2012}, {\em 419},~3426--3432,
	\href{http://xxx.lanl.gov/abs/1107.4831}{{\normalfont
			[arXiv:astro-ph.HE/1107.4831]}}.
	\newblock
	doi:{\changeurlcolor{black}\href{https://doi.org/10.1111/j.1365-2966.2011.19983.x}{\detokenize{10.1111/j.1365-2966.2011.19983.x}}}.
	
	\bibitem[{Bosch-Ramon} \em{et~al.}(2017){Bosch-Ramon}, {Barkov}, {Mignone}, and
	{Bordas}]{bbmb17}
	{Bosch-Ramon}, V.; {Barkov}, M.V.; {Mignone}, A.; {Bordas}, P.
	\newblock {HESS J0632+057: hydrodynamics and non-thermal emission}.
	\newblock {\em \mnras} {\bf 2017}, {\em 471},~L150--L154,
	\href{http://xxx.lanl.gov/abs/1708.00066}{{\normalfont
			[arXiv:astro-ph.HE/1708.00066]}}.
	\newblock
	doi:{\changeurlcolor{black}\href{https://doi.org/10.1093/mnrasl/slx124}{\detokenize{10.1093/mnrasl/slx124}}}.
	
	\bibitem[{Barkov} and {Bosch-Ramon}(2018)]{2018MNRAS.479.1320B}
	{Barkov}, M.V.; {Bosch-Ramon}, V.
	\newblock {A hydrodynamics-informed, radiation model for HESS J0632 + 057 from
		radio to gamma-rays}.
	\newblock {\em \mnras} {\bf 2018}, {\em 479},~1320--1326,
	\href{http://xxx.lanl.gov/abs/1806.05629}{{\normalfont
			[arXiv:astro-ph.HE/1806.05629]}}.
	\newblock
	doi:{\changeurlcolor{black}\href{https://doi.org/10.1093/mnras/sty1661}{\detokenize{10.1093/mnras/sty1661}}}.
	
	\bibitem[{Barkov} and {Bosch-Ramon}(2016)]{bb16}
	{Barkov}, M.V.; {Bosch-Ramon}, V.
	\newblock {The origin of the X-ray-emitting object moving away from PSR
		B1259-63}.
	\newblock {\em \mnras} {\bf 2016}, {\em 456},~L64--L68,
	\href{http://xxx.lanl.gov/abs/1510.07764}{{\normalfont
			[arXiv:astro-ph.HE/1510.07764]}}.
	\newblock
	doi:{\changeurlcolor{black}\href{https://doi.org/10.1093/mnrasl/slv171}{\detokenize{10.1093/mnrasl/slv171}}}.
	
	\bibitem[{Mignone} \em{et~al.}(2007){Mignone}, {Bodo}, {Massaglia}, {Matsakos},
	{Tesileanu}, {Zanni}, and {Ferrari}]{mbm07}
	{Mignone}, A.; {Bodo}, G.; {Massaglia}, S.; {Matsakos}, T.; {Tesileanu}, O.;
	{Zanni}, C.; {Ferrari}, A.
	\newblock {PLUTO: A Numerical Code for Computational Astrophysics}.
	\newblock {\em \apjs} {\bf 2007}, {\em 170},~228--242,
	\href{http://xxx.lanl.gov/abs/arXiv:astro-ph/0701854}{{\normalfont
			[arXiv:astro-ph/0701854]}}.
	\newblock
	doi:{\changeurlcolor{black}\href{https://doi.org/10.1086/513316}{\detokenize{10.1086/513316}}}.
	
	\bibitem[{Li}(2005)]{2005JCoPh.203..344L}
	{Li}, S.
	\newblock {An HLLC Riemann solver for magneto-hydrodynamics}.
	\newblock {\em Journal of Computational Physics} {\bf 2005}, {\em
		203},~344--357.
	\newblock
	doi:{\changeurlcolor{black}\href{https://doi.org/10.1016/j.jcp.2004.08.020}{\detokenize{10.1016/j.jcp.2004.08.020}}}.
	
	\bibitem[{Bosch-Ramon} \em{et~al.}(2012){Bosch-Ramon}, {Barkov}, {Khangulyan},
	and {Perucho}]{bbkp12}
	{Bosch-Ramon}, V.; {Barkov}, M.V.; {Khangulyan}, D.; {Perucho}, M.
	\newblock {Simulations of stellar/pulsar-wind interaction along one full
		orbit}.
	\newblock {\em \aap} {\bf 2012}, {\em 544},~A59,
	\href{http://xxx.lanl.gov/abs/1203.5528}{{\normalfont
			[arXiv:astro-ph.HE/1203.5528]}}.
	\newblock
	doi:{\changeurlcolor{black}\href{https://doi.org/10.1051/0004-6361/201219251}{\detokenize{10.1051/0004-6361/201219251}}}.
	
	\bibitem[{Aharonian} \em{et~al.}(2005){Aharonian}, {Akhperjanian}, {Aye}, and
	{et~al.}]{aha05}
	{Aharonian}, F.; {Akhperjanian}, A.G.; {Aye}, K.M.; {et~al.}.
	\newblock {Discovery of the binary pulsar PSR B1259-63 in very-high-energy
		gamma rays around periastron with HESS}.
	\newblock {\em \aap} {\bf 2005}, {\em 442},~1--10,
	\href{http://xxx.lanl.gov/abs/arXiv:astro-ph/0506280}{{\normalfont
			[arXiv:astro-ph/0506280]}}.
	\newblock
	doi:{\changeurlcolor{black}\href{https://doi.org/10.1051/0004-6361:20052983}{\detokenize{10.1051/0004-6361:20052983}}}.
	
	\bibitem[{Lyne} \em{et~al.}(2015){Lyne}, {Stappers}, {Keith}, {Ray}, {Kerr},
	{Camilo}, and {Johnson}]{lyn15}
	{Lyne}, A.G.; {Stappers}, B.W.; {Keith}, M.J.; {Ray}, P.S.; {Kerr}, M.;
	{Camilo}, F.; {Johnson}, T.J.
	\newblock {The binary nature of PSR J2032+4127}.
	\newblock {\em \mnras} {\bf 2015}, {\em 451},~581--587,
	\href{http://xxx.lanl.gov/abs/1502.01465}{{\normalfont
			[arXiv:astro-ph.HE/1502.01465]}}.
	\newblock
	doi:{\changeurlcolor{black}\href{https://doi.org/10.1093/mnras/stv236}{\detokenize{10.1093/mnras/stv236}}}.
	
	\bibitem[{Dubus} \em{et~al.}(2010){Dubus}, {Cerutti}, and
	{Henri}]{2010MNRAS.404L..55D}
	{Dubus}, G.; {Cerutti}, B.; {Henri}, G.
	\newblock {The relativistic jet of Cygnus X-3 in gamma-rays}.
	\newblock {\em \mnras} {\bf 2010}, {\em 404},~L55--L59,
	\href{http://xxx.lanl.gov/abs/1002.3888}{{\normalfont
			[arXiv:astro-ph.HE/1002.3888]}}.
	\newblock
	doi:{\changeurlcolor{black}\href{https://doi.org/10.1111/j.1745-3933.2010.00834.x}{\detokenize{10.1111/j.1745-3933.2010.00834.x}}}.
	
	\bibitem[{Zdziarski} \em{et~al.}(2018){Zdziarski}, {Malyshev}, {Dubus},
	{Pooley}, {Johnson}, {Frankowski}, {De Marco}, {Chernyakova}, and
	{Rao}]{2018MNRAS.479.4399Z}
	{Zdziarski}, A.A.; {Malyshev}, D.; {Dubus}, G.; {Pooley}, G.G.; {Johnson}, T.;
	{Frankowski}, A.; {De Marco}, B.; {Chernyakova}, M.; {Rao}, A.R.
	\newblock {A comprehensive study of high-energy gamma-ray and radio emission
		from Cyg X-3}.
	\newblock {\em \mnras} {\bf 2018}, {\em 479},~4399--4415,
	\href{http://xxx.lanl.gov/abs/1804.07460}{{\normalfont
			[arXiv:astro-ph.HE/1804.07460]}}.
	\newblock
	doi:{\changeurlcolor{black}\href{https://doi.org/10.1093/mnras/sty1618}{\detokenize{10.1093/mnras/sty1618}}}.
	
	\bibitem[{Sinitsyna} and {Sinitsyna}(2021)]{2021AN....342..337S}
	{Sinitsyna}, V.G.; {Sinitsyna}, V.Y.
	\newblock {Cyg X 3: A gamma ray binary}.
	\newblock {\em Astronomische Nachrichten} {\bf 2021}, {\em 342},~337--341.
	\newblock
	doi:{\changeurlcolor{black}\href{https://doi.org/10.1002/asna.202113930}{\detokenize{10.1002/asna.202113930}}}.
	
	\bibitem[{Massi} \em{et~al.}(2020){Massi}, {Chernyakova}, {Kraus}, {Malyshev},
	{Jaron}, {Kiehlmann}, {Dzib}, {Sharma}, {Migliari}, and
	{Readhead}]{2020MNRAS.498.3592M}
	{Massi}, M.; {Chernyakova}, M.; {Kraus}, A.; {Malyshev}, D.; {Jaron}, F.;
	{Kiehlmann}, S.; {Dzib}, S.A.; {Sharma}, R.; {Migliari}, S.; {Readhead},
	A.C.S.
	\newblock {Evidence for periodic accretion-ejection in LS I
		+61{\textdegree}303}.
	\newblock {\em \mnras} {\bf 2020}, {\em 498},~3592--3600,
	\href{http://xxx.lanl.gov/abs/2010.08598}{{\normalfont
			[arXiv:astro-ph.HE/2010.08598]}}.
	\newblock
	doi:{\changeurlcolor{black}\href{https://doi.org/10.1093/mnras/staa2623}{\detokenize{10.1093/mnras/staa2623}}}.
	
	\bibitem[{Barkov} and {Bosch-Ramon}(2022)]{brb21}
	{Barkov}, M.V.; {Bosch-Ramon}, V.
	\newblock {Relativistic hydrodynamical simulations of the effects of the
		stellar wind and the orbit on high-mass microquasar jets}.
	\newblock {\em \mnras} {\bf 2022}, {\em 510},~3479--3494,
	\href{http://xxx.lanl.gov/abs/2112.04202}{{\normalfont
			[arXiv:astro-ph.HE/2112.04202]}}.
	\newblock
	doi:{\changeurlcolor{black}\href{https://doi.org/10.1093/mnras/stab3609}{\detokenize{10.1093/mnras/stab3609}}}.
	
	\bibitem[{Bosch-Ramon} and {Barkov}(2011)]{bb11}
	{Bosch-Ramon}, V.; {Barkov}, M.V.
	\newblock {Large-scale flow dynamics and radiation in pulsar {$\gamma$}-ray
		binaries}.
	\newblock {\em \aap} {\bf 2011}, {\em 535},~A20,
	\href{http://xxx.lanl.gov/abs/1105.6236}{{\normalfont
			[arXiv:astro-ph.HE/1105.6236]}}.
	\newblock
	doi:{\changeurlcolor{black}\href{https://doi.org/10.1051/0004-6361/201117235}{\detokenize{10.1051/0004-6361/201117235}}}.
	
	\bibitem[{Bosch-Ramon}(2011)]{bos11}
	{Bosch-Ramon}, V.
	\newblock {Radio emission from high-mass binaries with non-accreting pulsars}.
	\newblock {\em ArXiv e-prints} {\bf 2011},
	\href{http://xxx.lanl.gov/abs/1103.2996}{{\normalfont
			[arXiv:astro-ph.HE/1103.2996]}}.
	
\end{thebibliography}

\end{document}